\documentstyle[hep93]{article}

\def\lessim{\lower0.6ex\hbox{$\,$\vbox{\offinterlineskip
\hbox{$<$}\vskip1pt\hbox{$\sim$}}$\,$}}

\def\grtsim{\lower0.6ex\hbox{$\,$\vbox{\offinterlineskip
\hbox{$>$}\vskip1pt\hbox{$\sim$}}$\,$}}

\def\up#1{\raise 1ex\hbox{\sevenrm#1}}

\begin{document}

\title{Gravitation, Experiment and 
Cosmology\footnote{}}\footnotetext{$^{*}$Based on lectures given at Les Houches
1992, SUSY-95 and Corfu 1995; to appear in the proceedings of the 5th Hellenic
School of Elementary Particle Physics.}

\author{Thibault Damour \\
        {\em Institut des Hautes Etudes Scientifiques, 91440 Bures sur Yvette,
France}\\
        {\em DARC, CNRS - Observatoire de Paris, 92195 Meudon Cedex,
France}
       }

\abstract{
The confrontation between general relativity (and its theoretically most
plausible deviations) and experimental or observational results is summarized.
Some discussion is devoted to the various methodologies used in confronting
theory and experiment. Both weak-field (solar system) and strong-field (binary
pulsar) tests are discussed in detail. A special discussion is devoted to the
cosmology of moduli fields, i.e. scalar fields having only
gravitational-strength couplings to matter.
}


\maketitle

\section{Introduction}\label{sec:intro}

During the present century, our knowledge of nature has been
drastically deepened by (among other things) the discovery of two new
interactions (weak and strong) and of a relativistic theory of the
gravitational interaction (general relativity). At present, much effort
is being directed towards uncovering a deeper level of description of
nature which would hopefully unify all the interactions. A minimum
requirement would be to unite the classical description of the
macroscopic spacetime structure provided by general relativity with the
quantum description of the microscopic world. From dimensional
considerations, it seems guaranteed that any theoretical description
encompassing both quantum theory (with its characteristic Planck
constant $\hbar$) and Einstein's theory (with its two constants $c$ and
$G$, Newton's constant) will look totally different from what we know on
the Planck length scale
\begin{equation}
\ell_P \equiv \left( {\hbar G\over c^3} \right)^{1/2} = 1.616 \times
10^{-33}\, \hbox{cm}\ , \label{eq:1.1a}
\end{equation}
corresponding to the energy scale
\begin{equation}
E_P \equiv \left( {\hbar c^5\over G} \right)^{1/2} =
1.221 \times 10^{19}\, 
{\rm GeV} \ . \label{eq:1.1b} 
\end{equation}
At present it is difficult to conceive of experiments probing directly
physical phenomena at the scale (\ref{eq:1.1a}). Except, maybe by observing
relics of the very early cosmological universe.  This line of thought will be
discussed below. There exists another route which might
inform us about the way gravity fits at a more fundamental level within the
scheme of all interactions.  Indeed, nearly all the attempts at unifying gravity
with the other interactions predict the existence of new long--range,
macroscopically coupled interactions appearing as ``partners" of gravitation. 
This is notably the case in string theory where gravity always appears
accompanied by a scalar field (the dilaton), and possibly by an antisymmetric
tensor field.  [Not to mention the many other fields that arise when
compactifying a higher--dimensional theory].  Our present theoretical
understanding of the generation of mass (i.e.  finite range) is much too poor to
allow one to make any prediction about the range of such possible partners.
Therefore, the possibility exists that the low--energy effective theory,
derived from a more unified theory, contains some other long--range
field mediating forces between macroscopic bodies.

In view of this possibility it is important to assess clearly what is
experimentally known about gravity, defined as the result of all the
unscreenable long--range interactions between macroscopic bodies. The
present lectures adopt, as systematically as possible, a field-theory
approach to gravitation and try to summarize what are the present
experimental constraints on any field--theoretical description of
gravity.  Beyond giving a catalogue of existing, and planned,
experiments, we try to extract the maximum theoretical information from
present data.  The two main questions that we address are:

(i) which elements of the present ``standard" description of gravity
(i.e. general relativity) have been really tested, and which have not~?
and,

(ii) what types of new fields with macroscopic couplings could have
naturally escaped detection so far, and what are the most promising
experiments to look for them~?

To complete the point of view adopted in these lectures the reader is
urged to consult the (still relevant) Les Houches 1963 lectures of
Dicke \cite{D64}, and the specialized book of Will \cite{W81}. 

In Section 7 below we shall turn our attention to cosmological constraints on
scalar fields having gravitational-strength couplings to matter.

\section{Methodologies for testing theories}\label{sec:2}

One can usefully distinguish two complementary approaches for testing
the experimental validity of any given theory. [Though we will apply
the following considerations only to gravitation theories, they have
a very general realm of validity]. These two approaches can be termed the
``phenomenological" one, and the ``theory-space" one, respectively. More
simply, they can be respectively characterized by the two verbs
``compare" and ``contrast". Before entering into the details of these
two approaches it may be useful to view the problem in purely logical
terms: Let $T$ denote a (scientific) theory, and $C$ some of its
(observable) consequences. It is well known that $T\Longrightarrow C$
is logically equivalent to (non $C$) $\Longrightarrow$ (non $T$). This
is the rationale for saying that experiments can ``falsify" a theory
and the basis of the phenomenological approach discussed below. In this
approach, experiments have mainly a ``negative" value, telling us
something about a theory only when it is ``wrong". On the other hand,
scientists would like to have a rationale for saying that they can
``verify" a theory. The only logical way of doing so seems to
consider the set of all possible theories say $\{T'\}$ and to
investigate which subset of $\{T'\}$, say $\{T_C\}$, implies the same
consequences $C$ as $T$. This is the basis of the theory--space
approach. This approach gives a more positive value to experiments
checking that $C$ holds: they tell us that the common features (if any)
of $\{T_C\}$ are ``true".

\subsection{Phenomenological approach (``compare")} \label{sec:2.1}

Let us assume that we dispose of a general ``kinematical" model,
containing several free parameters, say $\{ p_i^{\rm pheno}\}$, for
describing the structure and evolution of some physical system.  By
(least-squares) fitting this model to the actual observations of the
physical system, we can ``measure" the values of all the
phenomenological parameters:  $p^{\rm obs}_i \equiv (p_i^{\rm
pheno})_{\rm best-fit}$.  We can then {\it compare} the observed values
$p^{\rm obs}_i$ to any theoretical prediction concerning the parameters
$p_i$, as deduced from the current standard theory.  The final outcome
of this procedure is a set of yes-no questions
\begin{equation}
 p^{\rm obs}_i = p^{\rm theory}_i \ ? \label{eq:2.1}
\end{equation}
Actually, each observed value $p^{\rm obs}_i$ comes out of the fitting
process equipped with some error bar, say $\sigma^{\rm obs}_i$
(corresponding to some confidence level, and including both statistical
and systematic errors).  Therefore, the questions (\ref{eq:2.1})
should be phrased in probabilistic terms.  Moreover, as will be clear
from the examples below, the theory never completely predicts the
numerical values of all the $p_i$'s but gives them as functions of some
underlying theoretical parameters, $p_i^{\rm theory} = F^{\rm theory}_i
(\lambda^{\rm theory}_a)$.  When one disposes of more phenomenological
parameters than theory parameters, one can eliminate the latter and
express the $p_i$'s in terms of a subset of them, say $p^{\rm theory}_i
= f^{\rm theory}_i (p^{\rm theory}_a)$, where the index $a$ runs only
over a subset of the range of the index $i$.

An example will clarify the phenomenological approach. In the 18th and
19th centuries several scientists realized that, independently of
Newton's theory, it was always possible to represent the motion of the
solar system by modelling each planetary motion as a perturbed Keplerian
motion, with time--varying Keplerian parameters:  $a$, $e$, $\omega$,
$i$, $\Omega$, $P$, $T_0$.  Moreover each time-varying parameter in the
previous list could be decomposed in secular and short-period parts
according to $p(t) = p_0 + \dot p t + {1\over 2} \ddot p t^2 + \cdots +
\Sigma_n p_{\omega_n} \cos (\omega_n t + \varphi_n )$.  This means that,
(nearly) independently of any theory, one can represent the motion of
the solar system by a list of (constant) parameters, say $\{ a^{\rm
Mercury}_0,\dot a^{\rm Mercury},\ldots ,\omega^{\rm Mercury}_0, \dot
\omega^{\rm Mercury},\ldots, a^{\rm Venus}_0,\ldots \}$.  In particular,
the fit between that extended Keplerian model and the observations
yielded a certain value for the ``secular periastron advance" of
Mercury, say $\dot\omega^{\rm obs}_{\rm Mercury}$.  If we now assume a
particular theory of gravity, we can (in principle) compute the
theoretical value of $\dot \omega_{\rm Mercury}$ in terms of the other
parameters in the model [in the process we must use auxiliary relations
to eliminate some not directly observable theoretical parameters, like
the masses of the planets, in terms of the directly observable
parameters of the phenomenological approach].  Finally, we can compare
$\dot\omega^{\rm Newton's~theory}_{\rm Mercury}$ to $\dot\omega^{\rm
obs}$. As was discovered by Le Verrier in the middle of the 19th century
this comparison exhibits a serious (now more than 90 sigma) disagreement
between Newton's theory and observations. By contrast the prediction
for $\dot\omega_{\rm Mercury}$ within general relativity, say
$\dot\omega_{\rm Mercury}^{\rm GR}$, is in close agreement (within one
sigma) with the observed value.  Therefore, one usually concludes that
the Mercury-perihelion test is invalidating (or falsifying) Newton's
theory, but confirming (or verifying) Einstein's theory.  The problem
with this conclusion (besides the fact that it is based on only one
test) is that the pure phenomenological comparison theory/observations
is telling us nothing about which elements of the theory are being
tested.  Which part of the structure of general relativity have we
actually checked~?  and which parts have played no r\^ole in the test
and have therefore not been probed at all~?  Are there other theories
which pass also with success the same test~?  To answer such questions
one needs to shift from the phenomenological approach to another one
which takes more into account the various structures of the considered
theories.

\subsection{Theory-space approach (``contrast")} \label{sec:2.2}

The idea of this second approach is to embed one's currently preferred
theory within a continuous space of alternative theories.  It is well
known that our ability to distinguish color nuances is greatly increased
if we bring next to each other two different nuances to make a contrast
between them.  In the same way, past experience has shown that one can
(sometimes) better unravel the inner structures present in a theory if
one contrasts it to a theory which is similar but different in some way.
To use in practice a space of ``contrasting" theories, one needs a way
of charting it.  In the simplest case this will mean that we can
continuously label the contrasting theories by means of a finite set of
real parameters, say $\{\beta_a\}$.  [In more complicated cases the
labelling will need an infinite set of real parameters, or a
parametrization in terms of arbitrary functions].

Having, on the one hand, a charted space of contrasting theories
(together with the predictions they make) and, on the other hand, an
actual set of experimental data, we can ask which subset of theories
are in better agreement with experiment. A standard quantitative
criterion for measuring the agreement between a set of data, say
$\{x^{\rm obs}_n \}$ together with their estimated one sigma error bars
$\{\sigma^{\rm obs}_n\}$, and a corresponding set of theoretical
predictions $\{ x^{\rm theory}_n \}$, is to compute the $\chi^2$
(``goodness of fit") statistics.  In our case $\chi^2$, for given
experimental data, will be a continuous function of the $\beta_a$
parameters labelling the theories (and therefore their predictions):
\begin{equation}
\chi^2 (\beta_a) = \sum_{n\in {\rm data~set}}
\left( {x^{\rm obs}_n - x^{\rm theory}_n (\beta_a) \over
\sigma^{\rm obs}_n} \right)^2 \ . \label{eq:2.2}
\end{equation}
It is useful to imagine the function $\chi^2 (\beta_a)$ as defining a
hypersurface rising above the finite-dimensional space of theories.
For instance if there are only two theory labels $(\beta_1, \beta_2)$,
the theory-space can be plotted as a horizontal two-dimensional plane,
say $(\beta_1, \beta_2) \equiv (x,y)$, so that $z=\chi^2 (x,y)$ defines
a usual surface in the three-dimensional space $(x,y,z)$. The best
agreement between observations and theory corresponds to the lowest
values of $\chi^2$ (which is by definition positive). Therefore one is
interested in the minima of $\chi^2 (\beta_a)$, and their surroundings,
i.e. in the hollows of the surface $z=\chi^2 (x,y)$. More precisely a
convenient way of measuring quantitatively the likelihood for some
theories to be compatible with the observed data is to consider
successive horizontal slices of the $\chi^2$ hypersurface above a
minimum, or equivalently level contours of $\chi^2$  in the space of the
parameters $\beta_a$ when considering only what happens in theory space.
To each difference in level above a minimum, say $\Delta \chi^2 = \chi^2
- \chi^2_{\min}$, one can attribute a certain {\it confidence level}
(C.L.), which depends also on the number of fitted parameters, i.e.
in our case the number of theory parameters that we consider. For
example, when there is only one parameter $(\chi^2(\beta_1))$ the
condition $\Delta \chi^2 \leq 1$ defines a 68~\% confidence interval
around $\beta^{\min}_1$ (``one sigma level") and $\Delta \chi^2 \leq 4$
a 95~\% confidence interval (``two sigma"). For two parameters
$((\chi^2 (\beta_1,\beta_2))$ the 68~\% C.L. corresponds to the
two-dimensional region $\Delta \chi^2 \leq 2.3$ in the $\beta_1$,
$\beta_2$ plane and the 95~\% one to $\Delta\chi^2 \leq 6.2$. Actually,
the convenient link we just described between confidence levels and level
contours of the specific function $\chi^2(\beta_a)$ obtained by fitting
to one particular set of data (the one realized in an actual experiment)
is a simplification.  This simple link holds only in particular cases
(e.g.  uncorrelated Gaussian noise and linear dependence on the
$\beta$'s), or in the limit of large number of data points.  In the
general case one should consider the best-fit parameters, $\beta^{\min}_a$
(those minimizing $\chi^2$), as random variables inheriting their
probabilistic characteristics (distribution function in the space of the
$\beta_a$'s) from the ones of the noisy data $x^{\rm obs}_n$ supposed to
be an arbitrary sample selected from a random process with known
statistical characteristics.  In other words, a more rigourous analysis
of the confidence level regions in $\beta$ space would need to use
Monte-Carlo methods for generating fake sets of ``observed" data, and
would then study the distribution of the corresponding best-fit
$\beta$'s.

Summarizing, the theory-space approach (``contrast") associates to
each independent set of experimental data some confidence region in
theory space at, say, the 90~\% C.L.. This immediately raises the
following questions: does the collection of confidence regions
corresponding to the various data sets admit a non-empty intersection~?
[If not, that would mean either that none of the considered theories
is correct, or that some sets of data contain systematic errors].
And, if there exists a non-empty intersection what is its shape in
theory space, i.e. what are the common features of the theories that
pass the considered tests~? As we see from the last question, the
theory-space approach is giving us a handle on what theoretical
structures are being actually probed by some sets of observations.

Let us give an example of the use of the theory-space approach.
Eddington introduced in 1923 \cite{E22} the idea that, in the
quasi-stationary weak-field context of solar-system experiments, it was
possible to chart many possible relativistic theories of gravitation
(different from Einstein's) by means of two (weak-field) theory
parameters, $\beta$ and $\gamma$.  [This idea was later extended by
Nordtvedt, and Will, \cite{N68a,W71,WN72} who introduced new weak-field,
theory parameters: $\xi$, $\alpha_1$, $\alpha_2$, $\alpha_3$, $\zeta_1$,
$\zeta_2$, $\zeta_3$, $\zeta_4$.  We shall see also below how it has
been recently possible to extend the theory-space approach to the strong
gravitational field regime].  As will be discussed in detail below, the
Eddington $\gamma$ parameter measures the average spin content of the
fields mediating the gravitational interaction (i.e.  as we shall see
the velocity-dependent or magneticlike gravitational forces), while
$\beta$ parametrizes the cubic vertex of gravitational interaction
(3-body force).  By convention, general relativity corresponds to the
values $\beta = \gamma =1$.

Let us now reconsider within the theory-space approach the
Mercury-perihelion test. A relativistic theory with Eddington parameters
$\beta$ and $\gamma$, say $T(\beta, \gamma)$, predicts the following
value for the secular advance of the perihelion of a planet with
semi-major axis $a$, eccentricity $e$, and orbital period $P$
\begin{equation}
\dot\omega^{T(\beta,\gamma)} = \dot\omega^{\rm Newton} +
{6\pi GM_\odot \over c^2 a(1-e^2) P} \times {2+2\gamma -\beta\over 3}
\ , \label{eq:2.3}
\end{equation}
where $M_\odot$ is the mass of the sun and where $\dot\omega^{\rm Newton}$
denotes the Newtonian prediction, which is mainly due to planetary
perturbations if one separates out the effect of the Earth spin
precession, and assumes [to simplify the discussion] that the quadrupole
moment of the sun is small enough to contribute negligibly.  We then see
that the comparison between the observed, $\dot\omega^{\rm obs} \pm
\sigma^{\rm obs}$, and the predicted, $\dot\omega^{T(\beta,\gamma)}$,
values of the Mercury perihelion advance defines a certain confidence
strip in the Eddington theory plane.  Present data yield a 68~\% C.L.
strip approximately given by
\begin{equation}
{2+2\gamma -\beta\over 3} \ = \ 1.000 \pm 0.001 \ , \label{eq:2.4}
\end{equation}
if one assumes that the adimensionalized quadrupole moment of the sun
$J_2 \sim 2\times 10^{-7}$ \cite{S90}.  By contrast with the
phenomenological approach which led to a yes-no alternative (in the
present case:  ``yes, general relativity passes the test"), the result
(\ref{eq:2.4}) of the theory-space approach has a much more precise
information content, namely:  yes, the values $\beta = \gamma = 1$
(obtained in the weak-field limit of general relativity) are compatible
with the Mercury-perihelion data, but so are all the values of $\beta$
and $\gamma$ lying in the infinite strip (\ref{eq:2.4}) [e.g.  ($\beta
=5$, $\gamma=3$) or $(\beta =-1,$ $\gamma =0)$, etc\dots].  Many
different relativistic theories of gravitation can pass this test which
probes only a particular combination of velocity-dependent and nonlinear
effects.

This example exhibits the possibility that the $\chi^2$ hypersurface
corresponding to a set of experimental data has the form of a long,
flat valley. This shows the need to perform other experiments to find
out where, along this valley, stands the correct theory. For example,
the experiments concerning the deflection of light by the sun probe the
parameter $\gamma$ independently of $\beta$ and reduce the domain of
allowed theories to a small parallelogram around the point $\beta
=\gamma =1$. Actually, the two examples of ``classic tests" of general
relativity that we just gave are somewhat outdated and must be replaced
by other tests as we shall discuss below. [The Mercury-perihelion test
is inconclusive because we have no direct experimental measurement of
the quadrupole moment of the sun, and the light deflection test is superseded
by radio-wave deflection and gravitational time delay tests].

\section{Testing what ?} \label{sec:3}

The previous section has exemplified the usefulness of embedding our
currently favored standard model of the gravitational interaction, i.e.
general relativity, within a continuum of alternative models. The next
question that arises is: what are the natural extensions of general
relativity to consider~? To answer this question we need first to take
a close look at the structure of general relativity.

\subsection{The two structural elements of general relativity}
\label{sec:3.1}

Einstein's theory of gravitation rests on two basic postulates:

i) gravity is mediated only by a long-range symmetric tensor field,
$g_{\mu\nu}$;

ii) $g_{\mu\nu}$ couples universally to all other (fermionic and bosonic)
fields by replacing everywhere (in kinetic and interaction terms) the
flat Minkowski metric $f_{\mu\nu} = {\rm diag} (-1, +1, +1, +1)$ of
Special Relativity. [See the Appendix for our notation].

In technical terms these postulates mean that the total action reads
\begin{equation}
S_{\rm tot} = S_g [g_{\mu\nu}] + S_m [\psi_m, g_{\mu\nu}] \ ,
\label{eq:3.1}
\end{equation}
where the ``gravitational" action $S_g$ is a functional of $g_{\mu\nu}$
only (without any other long-range field, and without any preassigned
structure, like $f_{\mu\nu}$), and where the ``matter" action $S_m$ is
that of the current standard model of particle physics [$\psi_m$
denoting both the fermionic (``matter") fields and the bosonic
(``interaction") ones] in which one replaces everywhere the flat metric
$f_{\mu\nu}$ (and its associated flat connection) by the curved one
$g_{\mu\nu}$: $f_{\mu\nu} \to g_{\mu\nu}$, $\partial_\mu \to
\nabla_\mu$. [With the usual subtlety that one must also introduce a
``square root" of $g_{\mu\nu}$, i.e. a vierbein, for writing down the
fermionic terms; see the Appendix].

The replacement requirement $f_{\mu\nu} \to g_{\mu\nu}$ is unambiguous
for the (spin 1/2) fermions and the (spin 1) gauge fields, but leaves
open the possibility of introducing an arbitrary dimensionless parameter
in the coupling of scalar fields to gravity ($\xi \sqrt g R(g)
\varphi^\dagger \varphi)$. In the case of the Higgs scalar doublet this
ambiguity has only unobservably small consequences at macroscopic
distance scales.

Let us now turn our attention to the gravitational part of the action,
$S_g[g_{\mu\nu}]$.  Weyl \cite{W22} and Cartan \cite{C22} (see also
Ref.~\cite{L72}) proved that the most general form of the action leading
to second-order field equations in 4 dimensions was
\begin{equation}
S_g [g_{\mu\nu}] = {c^4\over 16\pi G} \int {d^4x\over c}\, \sqrt g
[R(g) - 2 \Lambda ] \ . \label{eq:3.2}
\end{equation}
The constants appearing in eq.~(\ref{eq:3.2}) are the velocity of light
$c$, the Newtonian gravitational constant $G$ and the cosmological
constant $\Lambda$ (with dimension [length]$^{-2}$).
Cosmological data indicate that the value of $\Lambda$ is at most of
a cosmological order of magnitude $(\Lambda < 3 (H_0/c)^2$ where
$H_0$ is the present value of the Hubble ``constant"). Such a small
value of $\Lambda$ makes its presence unobservable in all
non-cosmological gravitational experiments. When discussing the latter
experiments we shall consider that ``general relativity" means
eq.~(\ref{eq:3.2}) with $\Lambda =0$.

Another way of justifying eq.~(\ref{eq:3.2}), with $\Lambda = 0$,
as being the unique, consistent description of a long-range symmetric
tensor field in four dimensions is to follow the approach initiated by
Feynman \cite{F63}. There is a unique action describing the excitations
of a massless symmetric tensor field $h_{\mu\nu}$ propagating in a flat,
four-dimensional spacetime which is irreducible and ghost-free (no
negative energy excitations). With a suitable definition of $h_{\mu\nu}$,
this unique action reads
\begin{equation}
S_2[h] = {1\over 2} \int {d^4 x\over c} \left( h_{\mu\nu} - {1\over 2}
h f_{\mu\nu} \right)  [ \hbox{\rlap{$\sqcup$}$\sqcap$} h_{\mu\nu}
+ \partial_{\mu\nu} h
- \partial_{\alpha\mu} h^\alpha_\nu
- \partial_{\alpha\nu} h^\alpha_\mu ] \ , \label{eq:3.3}
\end{equation}
where $h\equiv h^\alpha_\alpha$, $\hbox{\rlap{$\sqcup$}$\sqcap$}
\equiv \partial^\alpha_\alpha$, the indices being raised by the flat
metric $f_{\mu\nu}$. Eq.~(\ref{eq:3.3}) admits the local gauge invariance
$h_{\mu\nu} \to h_{\mu\nu} +\partial_\mu \xi_\nu + \partial_\nu \xi_\mu$,
the presence of which ensures that only positive-energy excitations
propagate. The necessity of preserving the existence of a local gauge
invariance restricts very much the possibility of coupling $h_{\mu\nu}$
to other fields and to itself. Work by many authors has shown that there
is a unique (modulo field redefinitions) way of coupling $h_{\mu\nu}$ in
a consistent fashion \cite{W65,OP65,WW65,D70,BD74,FF79,W86}
[In other words there is a unique, consistent deformation of the linear
gauge invariance of massless spin 2 fields].  This unique answer is
equivalent to the expansion in powers of $\kappa$ of eq.~(\ref{eq:3.1})
with $g_{\mu\nu}= f_{\mu\nu} + \kappa h_{\mu\nu}$ (where $\kappa =
\sqrt{32\pi G/c^4}).$

It is important to keep in mind in the following the two basic
structural elements of general relativity. On the one hand, the coupling
of gravity to all the fields representing matter and its binding forces in
the Standard Model is described by a ``universal metric coupling",
\begin{equation}
S_m = S_{\rm Standard~Model} [\psi_m, g_{\mu\nu}] \ , \label{eq:3.4}
\end{equation}
and, on the other hand, the dynamics of the gravitational field itself
(propagation and self-interaction) is described by the Einstein-Hilbert
action (\ref{eq:3.2}). Correspondingly to these two elements of
structure, there will be experimental tests that probe the way gravity
couples to matter (in particular the universal features of that coupling)
and tests probing the structure and dynamics of the gravitational field
itself (e.g.  its spin and its range).

\subsection{New, macroscopic fields and couplings} \label{sec:3.2}

Alternative theories of gravitation are defined by introducing, besides
$g_{\mu\nu}$, new, long-range fields mediating the gravitational
interaction.  However, one should beware that, apart from tensor-scalar
theories, all the ``alternative gravitation theories" that have been
discussed in the specialized literature (notably Ref.~\cite{W81}) suffer
from various field-theory pathologies:  unboundedness from below of the
energy, negative-energy (ghost) excitations, algebraic inconsistencies
among the field equations, discontinuities in the degree-of-freedom
content, causality problems, etc.  The number of non-pathological field
theories that one can construct is actually very restricted.

In order to prevent any semantic confusion, let us emphasize that, in
these lectures, we mean by ``gravity" (or ``gravitational interaction")
the actual, observable interaction between macroscopic bodies which
extends over macroscopic ranges and cannot be screened by presently
known means.  With this definition, any field having a macroscopic range
(say $\lambda > 0.1$~mm), and coherent couplings to electrically neutral
bodies, will be said to participate in the gravitational interaction or,
for short, to be a gravitational field.  The aim of these lectures is to
summarize what is experimentally known about gravity, and to assess what
are the various gravitational fields compatible with the present
experimental evidence.

Besides the usual Einsteinian field $g_{\mu\nu}$, with its nearly
uniquely defined coupling to the matter of the Standard Model, there
is no theoretical shortage of fields that could contribute to mediating
gravity. [We consider only Bosonic fields; see \cite{F63} for a discussion
of the difficulties arising when using the multiple exchange of massless
Fermionic fields to generate a gravitational-like force].

First, there could be one, or several, {\it scalar fields}, say
$\varphi^1$, $\varphi^2$, \dots, $\varphi^n$. Scalar fields can exhibit
a rich variety of couplings to matter. To quote a few: scalar Yukawa
coupling to fermions, $g_S \varphi \bar \psi \psi$, pseudo-scalar Yukawa
coupling $g_P \varphi \bar \psi \gamma_5 \psi$, ``conformal" or
``metric" coupling $S_m [\psi_m, A^2 (\varphi) g_{\mu\nu}$] (which means
a universal coupling to the trace of the energy-momentum tensor, see
below), dilaton-like coupling to gauge fields, $\varphi {\rm Tr}(F^2)$,
axion-like coupling to gauge fields, $\varphi {\rm Tr} (FF^*),$ etc.
Moreover, scalar fields can have an infinite range or a finite one
without any restriction on their sources (contrarily to massless, gauge
fields whose couplings are strongly constrained by algebraic consistency
requirements, e.g.  $\partial_\nu F^{\mu\nu} = J^\mu \Longrightarrow
\partial_\mu J^\mu = 0$).  The existence of at least one sort of scalar
(by constrast to pseudo-scalar) coupling in the list above suffices to
generate a coherent interaction between (unpolarized) macroscopic
bodies.  Many theoretical models have suggested the existence in nature
of scalar fields:  dimensional reduction, extended supersymmetry,
dynamical solutions to the strong CP problem, the family problem, or the
problem of the appearance of particular mass scales, string theory, etc.

Second, there could exist (one or several) {\it vector fields}, $B_\mu$.
If a vector field is massive (i.e.  of finite range) its couplings to
matter are not restricted by any consistency requirement (at least at
the classical level).  If it is massless, or acquires a mass only
through spontaneous symmetry breaking, it must be coupled to a conserved
current.  In fact, there are several ``unused" conserved currents in the
Standard Model and it has been suggested that they could correspond to
new, macroscopic interactions \cite{LY,Fayet,S79,F86}.

Finally, let us mention the possible existence of an {\it antisymmetric
tensor field}, $B_{\mu\nu} = - B_{\nu\mu}$. This possibility was raised
long ago by Einstein and others \cite{E46}, and has been revived by
string theory \cite{SS74}.  As is expected for any gauge field, the
couplings of a massless $B_{\mu\nu}$ are severely restricted by
consistency requirements, thereby disqualifying the old ``unified"
theory of Einstein and its modern avatars \cite{DDM1}.  On the other
hand, a finite range $B_{\mu\nu}$ can exhibit a host of
phenomenologically interesting macroscopic couplings \cite{DDM2}.

Besides the scalar, vector and antisymmetric fields (of any order),
there are no bosonic fields which are known to have consistent couplings
with matter and Einsteinian gravity. For example, there is no known
way of coupling to gravity, in a fully consistent manner, a second
symmetric tensor field, be it massless or massive (see \cite{AD},
\cite{BD72} and references therein).

\section{Testing the coupling of matter to an external gravitational
field} \label{sec:4}

\subsection{Experimental consequences of universal metric coupling}
\label{sec:4.1}

We shall refer to the matter-gravity coupling (\ref{eq:3.4}), i.e.
\begin{equation}
S_m = S_{\rm Standard~Model} [\psi_m, g_{\mu\nu}] \ , \label{eq:4.1}
\end{equation}
as a ``universal metric coupling": all the different fields $\psi_m$
entering the Standard Model description of matter and its binding forces
feel the gravitational influence of the external world only through
their coupling to one and the same metric tensor $g_{\mu\nu}$. [We have
in mind here the case of a test system, of negligible self-gravity,
experiencing some background gravitational field described by
$g_{\mu\nu}$].  Note that, besides general relativity, there are
infinitely many examples of gravitation theories exhibiting a universal
metric coupling (they define the so-called class of metric theories of
gravity).  Indeed, the tensor $g_{\mu\nu}$ to which matter is coupled
needs not satisfy the dynamics derived from the Einstein-Hilbert action
(\ref{eq:3.2}), but could instead be algebraically constructed from
other fields having their own propagation properties.  The simplest
example of a non-Einsteinian metric theory is a tensor-scalar theory
where the $g_{\mu\nu}$ that couples to matter via eq.~(\ref{eq:4.1}) is
of the form
\begin{equation}
g_{\mu\nu} = A^2 (\varphi)\ g^*_{\mu\nu} \ , \label{eq:4.2}
\end{equation}
where $\varphi$ is a massless scalar field (with kinetic term
$\sqrt{g_*} g^{\mu\nu}_* \partial_\mu \varphi \partial_\nu \varphi)$,
$g^*_{\mu\nu}$ a massless spin 2 field (with kinetic term $\sqrt{g_*}
R(g_*)$), and $A(\varphi)$ an arbitrary function of $\varphi$.

Let us now recall a simple, but useful, general mathematical result
about (pseudo-) Riemannian spaces (due to Fermi and Cartan). We shall
phrase it for the case of a four-dimensional Lorentzian manifold
$(V,g)$ of signature $-+++$: Given any worldline ${\cal L}$ in $V$
({\it not} necessarily a geodesic), there always exist coordinate
systems $x^\mu$ (i.e. a map $\varphi$ from the abstract $V$ to $I\!\!R^4$)
such that the corresponding components of the metric satisfy all along
${\cal L}$ the following conditions
\begin{eqnarray}
[g_{\mu\nu} (x^\alpha)]_{\cal L} &=& f_{\mu\nu} \equiv {\rm diag}
(-1, +1, +1, +1) \ , \nonumber \\
\left[ {\partial g_{\mu\nu} (x^\alpha)\over \partial x^\lambda}
\right]_{\cal L} &=& 0 \ . \label{eq:4.3}
\end{eqnarray}
In other words (using Taylor's formula), the metric components
$g_{\mu\nu} (x^\lambda)$ in those special coordinate systems are equal,
all over a world tube enclosing ${\cal L}$, to the usual (constant)
components of a flat metric $f_{\mu\nu}$ modulo terms which are of
{\it second order} in the distance away from ${\cal L}$.  It is
easy to see that in the particular case where the abstract worldline
${\cal L}$ is assumed to be a geodesic its image in $I\!\!R^4$,
$L=\varphi ({\cal L})$, i.e.  its coordinate representation in one of
the ``good" coordinate systems $x^\mu$ satisfying eqs.~(\ref{eq:4.3}),
is a straight line $(x^\mu (s) = x^\mu (0) + su^\mu)$.  In the general
case, $L$ is a curved line in $I\!\!R^4$.

Let us now consider some physical system evolving according to
eq.~(\ref{eq:4.1}) in some given external gravitational field
$g_{\mu\nu}$, but isolated from the influence of any other external,
non-gravitational field. We assume that the gravitational field
generated by this system is everywhere negligible. Let us introduce a
world tube, say ${\cal T}$, of spatial radius $\sim d$, which encloses
completely the system at all times. We can apply the theorem
(\ref{eq:4.3}) within the world tube ${\cal T}$, i.e. define a special
coordinate system constructed along  some central world line ${\cal L}$
(to be identified later as the center-of-mass world line of the physical
system). In the limit where the physical system becomes very small with
respect to the characteristic scale of variation of $g_{\mu\nu}
(x^\lambda)$ we can neglect terms of order $d^2$, i.e. we can consider
that the spacetime metric is {\it flat} within ${\cal T}$:
$g_{\mu\nu} (x^\lambda) = f_{\mu\nu} + O(d^2) \simeq f_{\mu\nu}$. In
this approximation, the external gravitational field has been
{\it effaced} within ${\cal T}$, and we can conclude that the
system will evolve (when viewed in the good coordinates $x^\mu$)
as if it was an isolated system in special relativity. This means in
particular that the physical experiments taking place within the system
will exhibit no preferred directions in space (spatial isotropy), and
no preferred velocity states (boost invariance). The local time
evolution of the system will depend only on the values of the coupling
constants and mass scales that enter the usual Standard Model.
Moreover, the (special relativistic) center of mass of the system will
follow a straight line in the $x^\mu$ coordinates, corresponding to a
geodesic in the abstract curved spacetime $(V, g)$.

Then, by comparing the Lagrangian for time-like geodesics $[-m
(-g_{\mu\nu} (x^\lambda) \dot x^\mu \dot x^\nu)^{1/2}]$ with the
well-known Lagrangian of a test mass in Newtonian gravity
$\left[ {1\over 2} m\dot{\bf x}^2 + m U ({\bf x},t) \right]$ we conclude
that, when using space-time coordinates $(x^0=ct, x^i)$ adapted to the
Newtonian limit, the time-time component of the metric must be given by
\[ g_{00} (x) = -1 +2U (x)/c^2 + O(1/c^4) \ . \]
As is well-known, this result allows one to predict that, when
intercomparing by means of electromagnetic signals two, identically
constructed, clocks located at two different positions in a static,
external gravitational potential $U({\bf x})$, one should observe a
difference in clock rates given by
\begin{equation}
\left[ {\tau_1 \over \tau_2} \right]_i
= \left[ {\nu_2\over \nu_1} \right]_i
= 1 + {1\over c^2} [U({\bf x}_1) - U({\bf x}_2) ] \ . \label{eq:4.*}
\end{equation}
In eq.~(\ref{eq:4.*}) the subscript $i$ means ``when intercompared" by
receiving signals from both clocks at some common location which can
be the location of either clock or, in fact, an arbitrary point (at rest
with respect to the clocks).

Summarizing, the assumption (\ref{eq:4.1}) of universal metric coupling
is a very strong one which has the following observable consequences
for the physics of localized systems embedded in external gravitational
fields:

$C_1$:~{\it Constancy of the constants}: the outcome of local
non-gravitational experiments depends only on the values of the coupling
constants and mass scales entering the laws of special relativistic
physics.  [In particular, the cosmological time evolution of the universe
at large has no influence on local experiments].

$C_2$:~{\it Local Lorentz invariance}: local non-gravitational
experiments exhibit no preferred directions in spacetime [i.e. neither
spacelike ones (isotropy), nor timelike ones (boost invariance)].

$C_3$:~{\it Universality of free fall}: small, non self-gravitating
bodies, isolated from non-gravitational external forces, follow
geodesics of the external spacetime $(V, g)$.  [In particular, two test
bodies, initially next to each other and at rest with respect to each
other, fall in the same way in an external gravitational field,
independently of their mass and composition].

$C_4$:~{\it Universality of gravitational redshift}: when intercompared
by means of electromagnetic signals two identically constructed clocks
exhibit the difference in clock rate (or redshift) given by
eq.~(\ref{eq:4.*}) independently of the nature and constitution of the
clocks.

Note that $C_4$ yields a non trivial prediction of universality even
when considering two clocks at the same location $({\bf x}_1={\bf x}_2$).
However, this particular case is contained in the more general
consequence $C_1$.

The consequence $C_3$ is also referred to as the ``Weak Equivalence
Principle", and the whole set of consequences $C_1 - C_4$ is sometimes
called the ``Einstein Equivalence Principle" \cite{W81}.  In the present
lectures, as we wish to distinguish clearly experimental facts from
theoretical assumptions we will often shun the use of such expressions.
Note that the consequences $C_1 - C_4$ concern only systems with
negligible self gravitational fields.  Indeed, the universal metric
coupling (\ref{eq:4.1}) leaves open the possibility for self-gravitating
systems to feel the external universe in ways that violate the
consequences $C_1 - C_4$.  We give examples of such violations in the
following.  In fact, such violations are generic among metric theories
of gravity, and general relativity stands out as one of the very few
(probably only two \cite{DEF1}) theories for which the consequences $C_1
- C_4$ hold true even for self-gravitating systems (``Strong Equivalence
Principle").

\subsection{Non-metric couplings and their observational consequences}
\label{sec:4.2}

In the previous subsection we discussed the four main observational
consequences of the postulate that the matter-gravity interaction is
described by the universal metric coupling (\ref{eq:4.1}). This is
sufficient for conceiving experiments that will test the correctness
of the postulate (\ref{eq:4.1}). In other words, we are in position
to apply the ``phenomenological" methodology of Sec.~\ref{sec:2.1}
above (for instance by comparing the free fall acceleration of test
bodies). As was said there, it is however useful to go one step further,
namely to embed the metric couplings (\ref{eq:4.1}) within a larger
class of non-metric couplings. Indeed, knowing what type of violation
of the consequences $C_1 - C_4$ of Sec.~\ref{sec:4.1} can arise when one
changes the basic assumption (\ref{eq:4.1}) helps very much in planning
and interpreting experiments.

\subsubsection{Dilaton-like couplings} \label{sec:4.2.1}

Kaluza-Klein theories and string theory naturally introduce couplings
between scalar fields and gauge fields of the form
\begin{equation}
S_{\rm dil} = - {\beta \over 4} \int d^4 x\, \sqrt g\, \varphi \,
{\rm tr} (F^2) \ , \label{eq:4.4}
\end{equation}
where the trace is taken over the gauge indices of some Yang-Mills field
strength,
\[ F^a_{\mu\nu} = \partial_\mu A^a_\nu - \partial_\nu A^a_\mu
+ f^{abc} A^b_\mu A^c_\nu \ , \]
so that ${\rm tr}(F^2)$ denotes $g^{\alpha\mu}g^{\beta\nu}F^a_{\alpha\beta}
F^a_{\mu\nu}$.  [The gauge potential $A^a_\mu$ is geometrically
normalized, i.e.  it contains the gauge coupling constant $g$].  In a
$D$-dimensional spacetime the metric combination $\sqrt g g^{\alpha\mu}
g^{\beta\nu}$ appearing in (\ref{eq:4.4}) scales as $\lambda^{(D-4)/2}$
under a Weyl rescaling $g_{\mu\nu} \to \lambda g_{\mu\nu}$.  Therefore
this combination is Weyl-invariant in $D=4$.  This means that one cannot
then reabsorb the scalar field $\varphi$ in eq.~(\ref{eq:4.4}) by a
suitable conformal redefinition of the metric $g_{\mu\nu}$.  Therefore
the term (\ref{eq:4.4}) cannot be written in the purely metric form
(\ref{eq:4.1}).  It is an intrinsically non-metric coupling.

To investigate the observable consequences of this non-metric coupling
term we can note that eq.~(\ref{eq:4.4}), added to the bare kinetic
term of the Yang-Mills field $-{\rm tr}(F^2)/4g^2_0$, (if there is any),
means that the Yang-Mills field has a field-dependent effective coupling
constant,
\begin{equation}
{1\over g^2_{\rm eff} (\varphi)} = {1\over g^2_0} + \beta \varphi \ .
\label{eq:4.5}
\end{equation}
The result (\ref{eq:4.5}) has two types of consequences: i) it predicts
that the locally measured coupling ``constants" of the Standard Model
will depend on space and time (violation of the consequence $C_1$
above), and ii) it entails that bodies of different compositions will
fall differently in an external gravitational field (violation of
$C_3$). Moreover, the {\it spatial} dependence of the coupling
constants will clearly affect the intercomparison of clocks based on
different physical principles and will violate the consequence $C_4$
above. To show this let us consider the simple case where the only
coupling constant which varies is the electromagnetic one $\alpha$
(fine-structure constant). A clock based on a Bohr-like 
atomic transition counts
time in units proportional to $\alpha^{-2} \tau_e$ where $\tau_e =
\hbar /m_e c^2$ is the ``Compton time" associated to the electron,
while clocks based on fine or hyperfine atomic transitions
involve higher powers of $\alpha^{-1}$. By
contrast a clock based on the stability of a cavity counts time in units
proportional to a multiple of the Bohr radius divided by $c$, i.e. to
$\alpha^{-1} \tau_e$, which differs in the exponent of $\alpha$.

The amount of spatial and temporal variability of $g_{\rm eff}$ depends
on the mass of the scalar field $\varphi$, and on the presence or
absence of other couplings of $\varphi$ to matter besides (\ref{eq:4.4}).
For instance, if $\varphi$ is massless and couples through (\ref{eq:4.4})
to the $SU(3)_c$ Yang-Mills field, it will be generated macroscopically
through the fact that the mass of protons and neutrons (and thereby
that of all nuclei) is currently believed to be mainly made of gluon
field energy. The cosmological expansion of the universe will then
induce a slow time variation of $\varphi$, reflected in a corresponding
secular change of the strong coupling constant.

The existence of a close link between the constancy of the coupling
``constants" and the universality of free fall has been pointed out
by Dicke \cite{D64} (see \cite{W81} for further references).  To see the
necessity of a violation of the universality of free fall in presence of
a dilaton-like coupling it is sufficient to note that the classical
action describing the motion of a test particle, say an atom, reads (in
units where $c=1$)
\begin{equation}
S_m = - \int m\, ds \ , \label{eq:4.6a}
\end{equation}
where $ds =(-g_{\mu\nu} (x) dx^\mu dx^\nu)^{1/2}$ and where $m$ denotes
the total mass-energy of the atom. The latter mass-energy depends on the
effective values of the various gauge coupling constants, say $\alpha_i
= g^2_{{\rm eff}(i)} / 4\pi$ where $i=1,2,3$ labels the gauge groups
$U(1)$, $SU(2)$, $SU(3)$ respectively. The $\varphi$-dependence of the
$\alpha_i$'s entails a corresponding $\varphi$-dependence, and therefore
a spacetime dependence, of $m$:
\begin{equation}
m(x) = m[\varphi (x)] = m [\alpha_i \{ \varphi (x) \} ] \ .
\label{eq:4.6b}
\end{equation}

Varying the action (\ref{eq:4.6a}) yields the equation of motion
\begin{equation}
{d^2 x^\mu\over ds^2} + \Gamma^\mu_{\rho\sigma}\, {dx^\rho\over ds}\,
{dx^\sigma\over ds} \,=\, - \left[ g^{\mu\nu} + {dx^\mu\over ds}\,
{dx^\nu\over ds} \right]\, {\partial_\nu m\over m}\ , \label{eq:4.7}
\end{equation}
where $\Gamma^\mu_{\rho\sigma}$ denote the Christoffel symbols of
$g_{\mu\nu}$, and $\partial_\nu m\equiv \partial m/\partial x^\nu =
(\partial m/\partial\varphi) (\partial\varphi/\partial x^\nu)$ the
spacetime gradient of the mass-energy (\ref{eq:4.6b}). The right-hand
side of eq.~(\ref{eq:4.7}) gives, for, say, an atom starting from
rest in the gravitational field of the Earth, the additional term
$\delta {\bf a} = - \mbox{\boldmath{$\nabla$}} m/m =
-\mbox{\boldmath{$\nabla$}} \ln (m)$ (beyond the usual ${\bf g}$
associated with $g_{00}$) in the free fall acceleration of an atom.  If
we compare the free-fall accelerations of two different atoms, labelled
$A$ and $B$, we find the difference
\begin{equation}
{\bf a}_A -{\bf a}_B =- \mbox{\boldmath{$\nabla$}} (\ln\,m_A - \ln\,m_B) =
-\sum^3_{i=1} \left[ {\partial \ln\,m_A\over \partial \ln \alpha_i}\,
-\, {\partial\ln\,m_B \over \partial\ln\,\alpha_i}\right]
\mbox{\boldmath $\nabla$} \ln\,\alpha_i \ . \label{eq:4.8}
\end{equation}
Since different atoms have different field contributions to their
mass energy we expect the brackets in the right-hand-side of
eq.~(\ref{eq:4.8}) to differ from zero. 

\subsubsection{Multi-metric couplings, antisymmetric tensor couplings
and local Lorentz invariance} \label{sec:4.2.2}

The previous subsection has exemplified how dilaton-like scalar
couplings introduce violations of the consequences $C_1$, $C_3$ and $C_4$
discussed in \S~\ref{sec:4.1}. However, scalar couplings introduce
(in first approximation) no violations of the consequence $C_2$,
because the value of a scalar field is Lorentz invariant. One needs
to consider non-metric couplings involving vectors or tensors to
exhibit gravitational violations of local Lorentz invariance \cite{PD62},
\cite{D64}.

Let us first recall how ``isotropy of space" shows up in a simple
physical situation. Let us consider Schr\"odinger's equation for an
Hydrogen atom,
\begin{equation}
H_0\psi \equiv {-\hbar^2\over 2m} \Delta\psi - {e^2\over r} \psi =
E\psi \ . \label{eq:4.9}
\end{equation}
In this context, ``isotropy of space" means the invariance of
eq.~(\ref{eq:4.9}) under arbitrary rotations around the origin. This
spherical symmetry comes from the fact that $\Delta =\delta^{ij}
\partial_{ij}$ and $r=(\delta_{ij} x^i x^j)^{1/2}$ are both expressed
in terms of the same Euclidean metric $\delta_{ij}$. At a deeper level,
the latter property comes from the fact that the kinetic terms of the
electron field $(\bar\psi \gamma^\mu \partial_\mu\psi - m \bar\psi\psi)$
and of the electromagnetic field $(f^{\alpha\mu} f^{\beta\nu}
F_{\alpha\beta} F_{\mu\nu})$ involve the same flat spacetime metric
$(\gamma^\mu \gamma^\nu +\gamma^\nu \gamma^\mu = 2f^{\mu\nu})$.  This
coincidence in the propagation properties of the electron and
electromagnetic fields will be, by definition, preserved in the case of
universal metric coupling (\ref{eq:4.1}).  By contrast if, for some
reason, the coupling to gravity of $\psi$ and $F_{\mu\nu}$ introduces
two different spacetime metrics (say a``matter" metric $g_{\mu\nu}^m$
for $\psi$ and a ``field" metric $g^F_{\mu\nu}$ for $F_{\mu\nu}$) then
there will be observable violations of the ``isotropy of space".  In
first approximation it is enough to consider constant metric
coefficients.  Let us use coordinates (\`a la eq.~(\ref{eq:4.3}))
adapted to the matter metric, i.e.  such that $g^m_{\mu\nu} =
f_{\mu\nu}$ (so that in particular $g^m_{ij} = \delta_{ij}$ for the
spatial components).  In these coordinates, the field metric will, in
general, fail to have the Minkowskian form.  In particular, the spatial
components of the {\it conformal} field metric, say $\tilde
g^F_{\mu\nu} = -g^F_{\mu\nu} / g^F_{00}$ (which are the only quantities
that matter), will be of the general form $\tilde g^F_{ij} = \delta_{ij}
+h_{ij}$ with $h_{ij} \not= 0$.

In eq.~(\ref{eq:4.9}) $r$ will be replaced by $r_F =(\tilde g^F_{ij} x^i
x^j)^{1/2}$. Keeping only the terms linear in $h_{ij}$ (assumed to be
very small) leads to an Hamiltonian of the form $H_0 +H_1$ where the
unperturbed Hamiltonian $H_0$ is that given by eq.~(\ref{eq:4.9}), while
the perturbation reads
\begin{equation}
H_1 = {e^2\over 2}\, h_{ij}\, {x^ix^j\over r^3} \ . \label{eq:4.10}
\end{equation}
The (first-order) shifts in the energy levels of the atom are then
obtained by diagonalizing the projection of the operator $H_1$ in the
subspace spanned by some degenerate eigenstate of $H_0$. Indeed, the
spherical symmetry of $H_0$ implies that the unperturbed eigenvalues
are exactly degenerate with respect to the magnetic quantum number
$m$ (we do not consider here the accidental degeneracy of the $1/r$
potential). The perturbation $H_1$ associated with $h_{ij}$ will lift
the spherical-symmetry degeneracy. This gives an observational handle
on the violation of spatial isotropy induced by the assumption that the
electromagnetic field couples to a different metric than the electron.

In actual experiments, one considers nuclear energy levels, rather than
atomic ones, and experimental situations where the spherical symmetry
degeneracy has been already lifted, e.g. by interaction with an external
magnetic field. Generalizing the calculation above leads to energy
shifts in the $|I,M>$ state (where $I$ is the nuclear spin, and $M$ its
projection on the magnetic axis)
\begin{equation}
(E_1)_{I,M} = -{e^2\over 2}\, h_{ij} <I,M| \sum_{A<B} \,
{x^i_{AB} x^j_{AB} \over r^3_{AB}} |I,M> \label{eq:4.11}
\end{equation}
where the indices $A,B$ label the protons in the nucleus and $x^i_{AB}
\equiv x^i_A - x^i_B$.

Evidently, only the trace-free part of $h_{ij}$ will induce $M$-dependent
shifts.  For simplicity, we approximate the nuclear-structure matrix
elements appearing in the r.h.s.  of eq.~(\ref{eq:4.11}) in terms of
those of the electric quadrupole moment of the nucleus $\widehat Q^{ij}
= \Sigma_A e(x^i_A x^j_A - {1\over 3} {\bf x}_A^2 \delta^{ij})$ and of
some characteristic radius $R$:
\begin{equation}
(E_1)_{I,M} \sim - {(Z-1)e\over R^3} \left( h_{ij} -{1\over 3} h_{ss}
\delta_{ij} \right) <I,M| \widehat Q_{ij} |I,M> \ . \label{eq:4.12}
\end{equation}
The electric quadrupole moment operator can be expressed in terms of
the nuclear spin $\widehat {\bf I}$ (and of $Q =Q_{zz}$, its maximum
eigenvalue) as
\[  \widehat Q_{ij} = Q\, {3\over 2I(2I-1)}\, \left[ \widehat I_i
\widehat I_j + \widehat I_j \widehat I_i - {2\over 3}\,
\widehat {\bf I}^2 \delta_{ij} \right]    \ . \]
This yields an explicit expression for the $M$-dependence of the
anisotropic energy shifts
\begin{equation}
(E_1)_{I,M} \sim - {(Z-1)e Q \over R^3} \left( h_{ij} -{1\over 3} h_{ss}
\delta_{ij} \right) \widehat B^i \widehat B^j \,
{3M^2 -I(I+1)\over I(2I-1)}   \ , \label{eq:4.13}
\end{equation}
in which $\hat {\bf B}$ denotes a unit vector in the direction of the
external magnetic field (quantization axis). Experimental limits on the
presence of such terms will be discussed below.

Let us complete this subsection concerned with possible theoretical
origins for terms like eq.~(\ref{eq:4.10}) by mentioning how they could
be induced by certain couplings between gauge fields and a massive
antisymmetric tensor field.  Indeed, if gravity is mediated in part by a
(finite-range) antisymmetric tensor field $B_{\mu\nu}$, it could couple
to gauge fields via terms of the form
\begin{equation}
-{\alpha\over 8} \, {\rm tr}\, [(B_{\mu\nu} F^{\mu\nu})^2] \ .
\label{eq:4.14}
\end{equation}
[Note the necessity of considering a massive $B_{\mu\nu}$; the gauge
invariance of a massless one would forbid an algebraic coupling of the
form (\ref{eq:4.14}).] When considering electromagnetism, and separating
out the terms quadratic in the electric field $E^i =F^{0i}$, one finds
that eq.~(\ref{eq:4.14}) is equivalent to having introduced (as we did
above in an ad hoc manner) a different metric coupled to the electric
field: namely $g^F_{ij} = \delta_{ij} + h_{ij}$ with
\begin{equation}
h_{ij} = \alpha\, B_{0i} B_{0j} \ . \label{eq:4.15}
\end{equation}

One should  note also that the coupling (\ref{eq:4.14}) implies not
only a violation of the consequence $C_2$ [including evidently the local
boost invariance, the external $B_{\mu\nu}$ introducing preferred
spacetime directions] but also of $C_3$: the coupling of $B$ to
$F$-field energy will, like the dilaton coupling, introduce a violation
of the universality of free fall at some level.

\subsubsection{Other couplings of matter to scalar, vector or tensor
fields and their experimental consequences} \label{sec:4.2.3}

In the previous two subsections we selected some specific types of
non-metric couplings to exemplify clear cut violations of a subset of
the consequences $C_1-C_4$. In the present subsection we wish to show by
means of examples that most couplings one can think of, involving scalar,
vector or tensor fields, will entail a violation of at least one
of the consequences $C_1 -C_4$.

As soon as a scalar field has Yukawa couplings, $g_S\varphi\, \bar\psi
\psi$, to some of the Fermions that constitute ordinary matter it will be
generated macroscopically by the matter external to the test system
we are considering. Then the coupling of this external $\varphi$ to the
fermions constituting the test system will violate $C_3$. Indeed,
universality of free fall means a coupling to the total mass-energy
content of test bodies, while the Yukawa interaction we are considering
couples to a total scalar charge of a composite body of the form
\begin{equation}
C_S = \sum_i g^i_S <\bar \psi_i \psi_i > \label{eq:4.16}
\end{equation}
where the index $i$ labels the various fermions, and $<>$ the quantum
average corresponding to the state of the body. It seems clear that
no choice of the basic coupling constants $g^i_S$ will be able to ensure
the exact proportionality of $C_S$ to the total mass. Indeed, even if
one chooses the coupling constants to the quarks and leptons so that
the scalar charge of individual protons, neutrons and electrons coincides
with their respective mass, the presence of nuclear and electromagnetic
binding energies will prevent $C_S$ to be proportional to the mass for
nuclei and atoms.

Note in passing that a scalar having only pseudoscalar couplings
$(g_P\varphi\bar\psi\gamma_5 \psi$, $\varphi\varepsilon^{\mu\nu\rho\sigma}
F_{\mu\nu} F_{\rho\sigma},\ldots)$ would not contribute to observable
gravity because ordinary matter will not generate macroscopic sources
for such a field (one would need spin-polarized bodies, time dependent
magnetic field configurations,\dots)

Let us consider vector fields, $B_\mu$. Contrary to the case of scalar
fields, there is a big difference between massive (finite-range) and
massless (infinite-range) vector fields. Indeed, massless vector fields
admit a gauge invariance which restricts very much their possible
couplings. They can couple only to a conserved vector source. In other
words, they are generated by a conserved quantity, such as baryon
number, lepton number,\dots Although many more possibilities are open
in the case of massive vector fields, current theoretical lore favours
the case of initially massless (gauge) fields, even if they are to
acquire a mass via some spontaneous symmetry breaking mechanism. In that
case, the source has to be a conserved quantity, but the observable
range of the vector interaction can be finite. If we consider ordinary,
electrically neutral matter, it offers only two possible conserved
quantities: baryon number $B=N+Z$ and lepton number $L=Z$ (here $N$
denotes the number of neutrons and $Z$ the number of protons or
equivalently of electrons). [Evidently, more possibilities would be open
if we were to consider more exotic types of matter. This possibility
should be kept in mind when discussing the ``gravitational" effects
of dark matter]. Then the total vector charge of, say, an atom
can be written in terms of a coupling constant $g_V$ and a mixing
angle $\theta_5$ as \cite{ADR86}
\begin{equation}
C_V = g_V [\cos\,\theta_5 B + \sin\, \theta_5 L ] \ . \label{eq:4.17}
\end{equation}
Again, there is no way of choosing $\theta_5$ such that $C_V$ becomes
proportional to the total mass $M$ of the considered atom. The best
approximation is obtained by choosing $\theta_5 = 0$, i.e. coupling to
baryon number only. However, in that case nuclear binding energy makes
for a non proportionality between $B$ and $M$ at the $10^{-3}$ level
(when comparing a pair of atoms).

Let us consider an antisymmetric tensor field, say $B_{\mu\nu}$.  Both
extended supergravities and string theory naturally introduce a massless
$B_{\mu\nu}$ as a partner of $g_{\mu\nu}$.  The gauge invariance
$(B_{\mu\nu} + \partial_\mu \chi_\nu - \partial_\nu \chi_\mu)$ of such a
field restricts very much its couplings.  However, this leaves the
possibility of couplings of the form
\begin{equation}
- {1\over 2}\, f\, \varepsilon^{\alpha\lambda\mu\nu} J_\alpha
\partial_\lambda B_{\mu\nu} \ , \label{eq:4.18}
\end{equation}
where $J_\alpha$ is any macroscopic current, which does not need to
be conserved. However, if $B_{\mu\nu}$ stays massless, it has only one
(scalar) degree of freedom in four dimensions and the interaction
(\ref{eq:4.18}) amounts to coupling this scalar to $\partial_\alpha
J^\alpha$. A more interesting case arises when one assumes (without being
able to exhibit natural mechanisms for achieving it) that the initially
massless $B_{\mu\nu}$ acquires a non-zero mass.  Under this assumption,
$B_{\mu\nu}$ has the three degrees of freedom of a massive vector field,
and the interaction (\ref{eq:4.18}) is equivalent to coupling that vector
to $J_\alpha$. This offers an interesting alternative motivation for the
existence of finite-range vector interactions coupled to macroscopic
currents $J_\alpha$ \cite{DDM2}. Note that the term (\ref{eq:4.14})
written in the previous section assumed an initially massive $B_{\mu\nu}$.
This assumption opens the possibility of many more interactions with
interesting phenomenological consequences of which (\ref{eq:4.14}) is
just an example.  However, there is at present no theoretical motivation
for introducing such a fundamentally massive field [Not to mention the
fact that the non-perturbative consistency of the generalized
interactions considered in Ref.~\cite{DDM2} has not been proven].

Finally, note that a common feature of all the non-metric field
interactions considered above ($\varphi F^2$, $\varphi \bar\psi \psi$,
$B_\mu\bar\psi\gamma^\mu\psi$, $B_{\mu\nu} \varepsilon^{\mu\nu\alpha\beta}
\partial_\alpha (\bar\psi \gamma_\beta \psi),\ldots)$ is their failure
to produce, in the case of a composite body (say an atom), a coupling to
the total mass-energy $M$ of the body.  By contrast, the special
relativistic result $M=\int d^3x T^{00}/c^2$ with $T^{\mu\nu} =
(2c/\sqrt g) \delta S_{\rm matter}/\delta g_{\mu\nu}$ shows why any
field $B_{\ldots}$ which enters the matter action only by modifying the
spacetime metric $(g_{\mu\nu} = g_{\mu\nu} [B_{\ldots}] \to \delta
g_{\mu\nu} = C^{...}_{\mu\nu} \delta B_{...})$ couples to the total mass
of a body.  [We consider here bodies initially at rest, $\int d^3
xT^{0i} = 0$, and in stationary equilibrium, so that $\int d^3T^{ij} =0$
by the virial theorem, see below].  Metric coupling is the only known
way to generate a coupling exactly proportional to $M$, i.e.  one which
ensures the universality of free fall $(M {\bf a} = {\bf F}$ with ${\bf
F} \propto M)$.  In other words, the universality of free fall (or weak
equivalence principle) plays a leading r\^ole among the consequences
$C_1 - C_4$, and deserves to be tested with the utmost precision available.

\subsection{Experimental results on the coupling of matter to an external
gravitational field} \label{sec:4.3}

The observable consequences of a universal metric coupling listed in
Subsection \ref{sec:4.1} above naturally lend themselves to
high-precision, null tests.

Many sorts of data (from spectral lines in distant galaxies to
measurement of solar-system isotopic abundances) have been used to set
limits on a possible time variation of the basic coupling constants
of the Standard Model \cite{Dyson}. For a recent laboratory test of a possible
variation of the fine-structure constant see \cite{PTM95}. The discovery of the
``Oklo Natural Reactor", a place in Gabon, Africa where sustained $U^{235}$
fission reactions occurred by themselves two billion years ago, gave data
that led to tightened limits on many constants \cite{S76,W81}.  In particular
ref.\cite{S76} quotes for the time variation of the electromagnetic and weak
(Fermi) coupling constants 
\begin{eqnarray}
|\dot \alpha / \alpha| &<& 5 \times 10^{-18} {\rm yr}^{-1}\ , \nonumber\\
|\dot G_F / G_F | &<&  10^{-12} {\rm yr}^{-1}\ .
\label{eq:4.19}
\end{eqnarray}
See, however, the global analysis of ref. \cite{33a} which leads to 
more conservative limits: e.g. $|\dot \alpha / \alpha| < 10^{-15}
{\rm yr}^{-1}$. Improving on previous (already very precise) results, recent
experiments \cite{HD} have obtained extremely tight limits on any possible space
anisotropy in nuclear energy levels. These experiments look for
time-dependent quadrupolar shifts of the (Zeeman) energy levels of
nuclei with spin~$ >1/2$ (in practice $I =3/2)$. In terms of the
expression (\ref{eq:4.13}) above this means essentially putting limits
on $\tilde h_{ij} \equiv h_{ij} -{1\over 3} h_{ss} \delta_{ij}$,
assuming that $\tilde h_{ij}$ remains fixed in a locally inertial
coordinate system, while $\widehat B^i$ (direction of the magnetic field
produced in the laboratory) rotates with the Earth. The best limits
so obtained are of the impressive order
\begin{equation}
\left| h_{ij} - {1\over 3} h_{ss} \delta_{ij} \right| \lessim 10^{-27}
 \ . \label{eq:4.20}
\end{equation}
Note that even if the conformal field metric $\tilde g_{\mu\nu}^F =
-g^F_{\mu\nu} / g^F_{00}$ introduced above happens to be isotropic in
some preferred frame (maybe some mean rest frame of the universe), say
$[\tilde g^F_{ij}]_{\rm preferred~frame} = (1+\epsilon) \delta_{ij}$,
the $\tilde h_{ij}$ entering Earth-based experiments will have an
anisotropic contribution $\sim \epsilon v^i v^j/c^2$ due to the motion
of the Earth (with velocity $v^i$) with respect to the preferred frame.
As one expects $v/c \sim 10^{-3}$ (both from our Galactic motion and our
motion with respect to the cosmic microwave background) the excellent
limit (\ref{eq:4.20}) yields the still very impressive $|\epsilon| \lessim
10^{-21}$.  One should however keep in mind the assumption (used in
setting the limit (\ref{eq:4.20})) that the source of anisotropy is
external to the rotating Earth.  It seems to me that (because of the
possible presence of ill-calibrated DC effects) the experiments performed
up to now do not put any interesting limits on an Earth-generated
``anisotropy of space", as e.g.  would be the case for the term
(\ref{eq:4.14}) if $B_{\mu\nu}$ had a finite range $\lessim$ the Earth
radius.

The universality of free fall has been tested by many high-precision
experiments (Bessel, E\"otv\"os, Renner, Dicke,\dots).  Actually most
experiments do not let the test masses fall but compare the forces needed
to hold them in place when submitted to the gravitational influence of
an external source (apparent gravitational forces in an Earth-based
frame).  Most modern experiments have used a torsion balance, i.e.  a
thin wire holding (in its simplest version) a rod at the extremities of which
are suspended two different bodies.  This apparatus measures the non parallelism
of the apparent gravitational forces acting on the two bodies.  Depending upon
the way the experiment is set and/or analyzed the results probe various
types of violations of the universality of free fall.  For instance the
Princeton experiment \cite{RKD,D64} looked for effects linked to the
apparent motion of the Sun.  This means that it was probing only fields
with range greater than or equal to the distance to the Sun.  In 1986,
hints of apparent violations of both the universality of free fall (in
the residuals of E\"otv\"os' experiments) and the inverse-square law (in
mine data) were presented as evidence for the existence of an
intermediate-range $(\lambda \sim 100$~m) force coupled to baryon number
(when considering non strange matter) \cite{F86}.  This suggestion has
spurred many new experiments, especially ones testing for possible
intermediate-range violations of the universality of free fall.  See
Ref.~\cite{A91} for a review of the experimental situation and a
detailed assessment of the constraints on the intensity, mixing angle
$\theta_5$ (see eq.~(\ref{eq:4.17})) and range of any new macroscopic
force.  Let us only quote here a sample of the present experimental
constraints on the fractional intensity $\tilde \alpha$, with respect to
gravity, of a force coupled to baryon number [i.e.  $\tilde \alpha
\equiv -g^2_V /4\pi G u^2$ for a vector interaction (\ref{eq:4.17}) with
$\theta_5 =0$ and $u\equiv 1$ atomic mass unit]:
\begin{eqnarray}
|\tilde \alpha| \lessim 10^{-3} \quad && \hbox{for}\quad
\lambda = 1\ {\rm m} \ , \nonumber \\
|\tilde \alpha| \lessim 2 \times 10^{-6} \quad && \hbox{for}\quad
\lambda = 1\ {\rm km} \ , \nonumber \\
|\tilde \alpha| \lessim 4 \times 10^{-9} \quad && \hbox{for}\quad
\lambda \geq  10 000\ {\rm km} \ . \label{eq:4.21}
\end{eqnarray}

If one considers infinite-range interactions, the direct
phenomenological limit on a possible differential free-fall acceleration
between two bodies is at the level \cite{A90} \cite{37a}
\begin{equation}
|{\bf a}_A - {\bf a}_B| / |{\bf a}| \lessim 3 \times 10^{-12} \ . \label{eq:4.22}
\end{equation}
One should also note that the most recent analyses of Lunar Laser Ranging
data (see section 5.3  below) find that the fractional difference in
gravitational acceleration toward the Sun between the (silica-dominated) Moon
and the  (iron-dominated) Earth is $ \lessim 10^{-12}$.

Finally, many experimental tests of the universality of the gravitational
redshift have been performed.  In the 1960's high-precision experiments,
making use of the M\"ossbauer effect, verified that the gravitational
redshift of a gamma ray line over a 22m difference in altitude was given
by eq.~(\ref{eq:4.*}) with 1~\% precision \cite{PRS}.  Other experiments
have used spectral lines in the Sun's gravitational field, stable clocks
transported on aircraft, rockets, satellites and spacecrafts, or have
compared Earth-bound clocks with the natural clock defined by the highly
stable millisecond pulsar PSR~1937+21 (For references see Ref.~\cite{W81}).
Some null redshift experiments $[{\bf x}_1={\bf x}_2$ in eq.(\ref{eq:4.*})]
have also been performed.  The most precise test to date of
eq.~(\ref{eq:4.*}) achieved a fractional accuracy on the gravitational
redshift $\simeq 2\times 10^{-4}$ \cite{VL}.  It consisted of flying a
hydrogen-maser clock on a rocket to an altitude $\sim$~10~000~km while
continuously comparing it to a similar clock on the ground.

\subsection{Theoretical conclusions about the coupling of matter to an
external gravitational field} \label{sec:4.4}

As we have summarized above, the main observable consequences of the
postulate (\ref{eq:4.1}) of universal metric coupling have been
verified with high precision by all existing experiments. Within the
presently achieved experimental resolution of many dedicated
experiments, there are no observational hints of violations of the
consequences $C_1 - C_4$. On the other hand, subsection \ref{sec:4.2}
above has shown, by way of examples, that all non-metric couplings
that suggest themselves within the present framework of theoretical
physics generically lead to violations of one or several of the
consequences $C_1 - C_4$. Therefore the simplest interpretation of the
present experimental situation is that the coupling of matter to an
external gravitational field is exactly of the metric form (\ref{eq:4.1}).

This conclusion should not however be interpreted as being final.  Let
us indeed examine critically the theoretical weight of the tests
reviewed in the previous subsection.  The most impressive experimental
limit is eq.~(\ref{eq:4.20}).  However, no really natural couplings
violating the local isotropy of space have been proposed [we exhibited
(\ref{eq:4.14}) as an example of field couplings violating $C_2$, but it
is rather ad hoc and assumes a massive antisymmetric tensor field to
start with].  The second most impressive observational limit is
eq.~(\ref{eq:4.22}).  However, as written down in the first
eq.~(\ref{eq:4.21}), the data behind (\ref{eq:4.22}) allow, e.g., for a
new field, with range $\lambda = 1$m, coupled to baryon number with
strength which can be as large as $10^{-3}$ times that of gravity.
There exist several models in which factors $\lessim 10^{-3}$ appear
naturally.  For instance the old suggestion \cite{S79} of a vector
partner of $g_{\mu\nu}$ coupled to the (PCAC) mass current of the
quarks, $m_u \bar u\gamma^\mu u+ m_d \bar d\gamma^\mu d+\cdots$,
generates a force between macroscopic bodies coupled (approximately) to
the combination $B-0.17~L$ with strength $[(m_u +2m_d)/m_N]^2 \simeq
3\times 10^{-4}$ weaker than gravity.  Even the existence of
infinite-range (massless) fields should not be dismissed.  The
couplings of such fields are tightly constrained by eqs.~(\ref{eq:4.19})
[where the first limit is $\sim 5\times 10^{-8}$ smaller than the Hubble
rate] and eq.~(\ref{eq:4.22}). It has been recently pointed out \cite{DN92,40a}
that such small coupling  strengths might be natural consequences of the
cosmological  evolution. In particular, ref. \cite{40a} suggests that the dilaton
(or one of the moduli fields) of string theory might exist in the 
low-energy world today as a weakly coupled massless field entailing 
very small violations of the consequences $C_1-C_4$.

In view of these possibilities, it is important to continue improving
the precision of the experimental tests of the consequences $C_1-C_4$.
In particular, let us mention the  project of a Satellite Test of
the Equivalence Principle \cite{STEP} (nicknamed STEP, and considered by ESA, 
NASA and CNES) which aims at probing the universality of free fall of pairs of
test masses orbiting the Earth at the impressive level $\delta a/a \sim
10^{-17}$.  Let us also note that there are plans for flying very stable clocks
near the Sun; the aim being to improve the testing of the gravitational redshift
down to the $10^{-6}$ fractional level, i.e.  the level where second-order
effects $\propto (U_{\rm sun} /c^2)^2$ enter eq.~(\ref{eq:4.*}) (Vessot).
See Ref.~\cite{M93} for a recent survey of these, and other, projects
in experimental gravity.

\section{Testing the newtonian and post-Newtonian limits of metric
theories of gravity}\label{sec:5}

\subsection{What are the most natural metric alternatives to Einstein's
theory ?}\label{sec:5.1}

In the rest of these lectures we shall adopt the provisional conclusion
of the previous section, namely that gravity couples to matter in the
purely metric way (\ref{eq:4.1}). This conclusion seems to leave open many
possibilities for alternative, non Einsteinian, theories of gravity.
Indeed, the physical metric tensor $g_{\mu\nu}$ through which matter
interacts with external gravity can still be an arbitrary function of
many different fields
\begin{equation}
g_{\mu\nu} = g_{\mu\nu} [ g^*_{\mu\nu},\, \varphi,\, B_\mu,\,
B_{\mu\nu},\ldots ] \ , \label{eq:5.1a}
\end{equation}
for instance
\begin{equation}
g_{\mu\nu} = A^2(\varphi) [g^*_{\mu\nu} + a_1 B_\mu B_\nu
+ a_2 g^*_{\mu\nu} g^{\rho\sigma}_* B_\rho B_\sigma
+ a_3 g^{\rho\sigma}_* B_{\mu\rho} B_{\nu\sigma} + \cdots ] \ .
\label{eq:5.1b}
\end{equation}
[We do not include field derivatives in eq.~(\ref{eq:5.1b}), e.g.
$\partial_\mu \varphi \partial_\nu \varphi$, because they induce serious
causality problems.] However, the appearance, besides a basic tensor
field $g^*_{\mu\nu}$ and one or several scalar fields $\varphi$, of
vector fields antisymmetric tensor fields, etc\dots in eq.~(\ref{eq:5.1a}) seems
theoretically improbable for the following reasons.  First, the non
gauge-invariance of the combinations $B_\mu B_\nu$ and $B_{\mu\sigma}
B_{\nu\rho}$ implies (if one wishes to avoid the presence of
negative-energy excitations) that the fields $B_\mu$ and $B_{\mu\nu}$
must have from the beginning a non-zero mass (or finite range), i.e.
more precisely that their kinetic terms must be of the form $-{1\over 4}
F^2_{\mu\nu}-{1\over 2} m^2 B^2_\mu$ and $-{1\over 12}H^2_{\lambda\mu\nu}
-{1\over 4}m^2 B^2_{\mu\nu}$ respectively, where $F_{\mu\nu} \equiv
\partial_\mu B_\nu -\partial_\nu B_\mu$, $H_{\lambda\mu\nu}=
\partial_\lambda B_{\mu\nu} + \partial_\mu B_{\nu\lambda} +\partial_\nu
B_{\lambda\mu}$.  [It is well known that all other forms for the kinetic
terms lead to ghost excitations].  This would mean the presence at a
fundamental level of the theory of a particular length scale $\lambda
=1/m$.  It does not seem very plausible that such a fundamental length
scale happens to be of macroscopic magnitude, as is necessary for it to
be relevant to the topic of these lectures [if $\lambda \sim \ell_P$,
eq.~(\ref{eq:1.1a}), there will be no observable consequences of the presence of
such fields].  A second reason which does not favor the existence of
fields $B_\mu$ and $B_{\mu\nu}$ coupling to matter only through
eq.~(\ref{eq:5.1a}) is that such fields would exhibit no {\it linear} couplings
to matter.  Their source $\delta S_{\rm matter} /\delta B = (\delta S_m /
\delta g_{\mu\nu}) (\delta g_{\mu\nu} /\delta B)$ is (at least) linear
in $B$, so that an everywhere vanishing $B$ field is an exact solution
of the $B$ field equations.  The only way the $B$ fields can couple to
local matter is through the presence of a cosmological $B$ background,
generated by putting suitable boundary conditions at the Big Bang.
Thirdly, the quadratic couplings of the $B$ fields to matter, e.g.
$a_1T^{\mu\nu} B_\mu B_\nu$ modify the mass terms in the action, and it
remains to be proven that these modifications preserve the consistency
of the theory.  Finally, though the ellipsis in eq.~(\ref{eq:5.1a})
could stand for other types of tensors (like a second symmetric tensor
field), we have seen above that it seems very difficult to introduce
such fields in a consistent way [i.e. free of algebraic inconsistencies,
discontinuities in the degree-of-freedom content, causality problems,
negative-energy excitations, etc\dots].

In conclusion, the most natural metric theories of gravity are expected
to contain only one symmetric tensor field, $g^*_{\mu\nu}$, and one
or several (massive or massless) scalar fields, $\varphi^a$,
$a=1,\ldots n$, and to couple to the Standard Model of matter via a
physical metric of the form
\begin{equation}
g_{\mu\nu} = A^2 (\varphi^1,\ldots, \varphi^n) g^*_{\mu\nu}
\label{eq:5.2}
\end{equation}
where $A(\varphi^a)$ is some arbitrary coupling function.
Note that if we require from the beginning to have only massless fields
the drastic consistency constraints on the couplings of gauge fields
(see e.g.  Refs.~\cite{F63,W65,OP65,WW65,D70,BD74,FF79,W86} and
\cite{DDM1,DDM2,AD,BD72}) force one to consider only tensor-multi-scalar
theories. [We do not consider here massless Fermion fields].

Finally, let us give a simple physical argument (which is not really
independent of the consistency ones given above) which shows clearly why
tensor-scalar theories are preferred when one assumes that consequences
$C_1-C_4$ are exactly satisfied.  In fact, let us start only from $C_3$,
the universality of free fall.  The usual reasoning $({\bf F}_A = M_A
{\bf a}_A$ with ${\bf a}_A = {\bf g}$ independent from which body $A$ is
considered) shows that gravity couples exactly to mass $({\bf F}_A \propto
M_A)$.  On the other hand, Special Relativity tells us that the mass of
a body (in stationary inner equilibrium) can be written either as
\begin{equation}
M = {1\over c^2} \int d^3 x\, T^{00} \ , \label{eq:5.3}
\end{equation}
or
\begin{equation}
M = {1\over c^2} \int d^3 x\,( T^{00} - T^{ss}) \ , \label{eq:5.4}
\end{equation}
where $T^{\mu\nu}$ denotes the total stress-energy tensor (including
matter and field contributions). Indeed, the second form is a consequence
of the virial theorem
\begin{equation}
\int d^3 x\, T^{ij} = {1\over 2}\, \partial^2_0 \int d^3 x\,
T^{00} x^i x^j \label{eq:5.5}
\end{equation}
(which follows directly from the conservation laws $\partial_\nu
T^{\mu\nu} = 0$). Eq.~(\ref{eq:5.5}) shows that the integrated stresses,
$\int d^3x T^{ij}$, and in particular their trace, vanish for a body
in stationary state.

Now, the first form (\ref{eq:5.3}) of the ``gravitational charge" suggests
a coupling to a massless spin-2 field, $h_{\mu\nu} T^{\mu\nu}$ (the
consistency of the coupling being ensured by the fact that $T^{\mu\nu}$
is conserved), while the form (\ref{eq:5.4}) suggests a scalar coupling
$\varphi T^\mu_\mu$.  At the linearized level, we thereby expect an
interaction of the general form
\begin{equation}
S_{\rm interaction} = {1\over 2} \int {d^4x\over c} (h_{\mu\nu} +
2\alpha_a \varphi^a f_{\mu\nu}) T^{\mu\nu}\ , \label{eq:5.6}
\end{equation}
where $f_{\mu\nu}$ denotes as above the flat metric and where the index
$a$ labelling the various possible scalar fields is summed over.
Remembering that $T^{\mu\nu}$ is the functional derivative of the matter
action with respect to the metric,
\begin{equation}
\delta S_m = {1\over 2} \int d^4 x \, \sqrt g\, T^{\mu\nu}
\delta g_{\mu\nu} \ , \label{eq:5.7}
\end{equation}
we conclude that, at the linearized level, eq.~(\ref{eq:5.6}) is telling
us that the gravitational couplings of matter is described by replacing
in the matter action the flat metric $f_{\mu\nu}$ by
\begin{equation}
g^{\rm linearized}_{\mu\nu} = f_{\mu\nu} + h_{\mu\nu} + 2\alpha_a
\varphi^a f_{\mu\nu} \simeq (1+\alpha_a \varphi^a)^2
[f_{\mu\nu} + h_{\mu\nu}] \ , \label{eq:5.8}
\end{equation}
where $h_{\mu\nu}$ is a massless spin-2 field and $\{ \varphi^a \}$ a
collection of (massless or massive) scalar fields. The result
(\ref{eq:5.8}) is nothing but the linearized version of eq.~(\ref{eq:5.2}).
The coupling coefficients $\alpha_a$ measuring the relative weight of
scalars with respect to the spin-2 field in the linearized gravitational
interaction are just the logarithmic gradients of the coupling function
$A(\varphi^a)$ of eq.~(\ref{eq:5.2}),
\begin{equation}
\alpha_a = {\partial\, \ln\, A(\varphi) \over \partial
\varphi^a}\ .
\label{eq:5.9}
\end{equation}
After differentiation, the r.h.s. of eq.~(\ref{eq:5.9}) is to be
evaluated at the background (or vacuum-expectation) values of the scalar
fields (e.g. $\varphi^a_0 = 0$), while the constant conformal factor
$A(\varphi_0)$ must be transformed away by rescaling the coordinates.
[More about this below].

\subsection{The Newtonian limit of tensor-multi-scalar theories and its
experimental tests} \label{sec:5.2}

Let us commence by defining in full detail the most general class of
tensor-multi-scalar theories. The total action reads
\begin{equation}
S_{\rm tot} = S_{g_*} + S_\varphi + S_m \ , \label{eq:5.10}
\end{equation}
with
\begin{eqnarray}
S_{g_*} &=& {c^4\over 4\pi G_*} \int {d^4x\over c}\,\sqrt g_*
\left[ {1\over 4}\, R_* \right] \ , \label{eq:5.11a}\\
S_\varphi &=& -{c^4\over 4\pi G_*} \int {d^4x\over c}\,\sqrt g_*
\left[ {1\over 2}\, g_*^{\mu\nu} \gamma_{ab} (\varphi^c)
\partial_\mu \varphi^a \partial_\nu \varphi^b +
B (\varphi^a) \right] \ , \label{eq:5.11b}
\end{eqnarray}
and
\begin{equation}
S_m = S_{\rm Standard~Model} [\psi_m,\, g_{\mu\nu}] \ , \label{eq:5.11c}
\end{equation}
in which the physical (or ``Jordan-Fierz") metric $g_{\mu\nu}$ directly
coupled to matter is related to the ``Einstein" one $g_{\mu\nu}^*$
appearing in the Einstein-Hilbert action (\ref{eq:5.11a}) (where $R_*$
denotes the Ricci scalar of $g^*_{\mu\nu}$) by a scalar-field dependent
conformal factor
\begin{equation}
g_{\mu\nu} = A^2 (\varphi^a) g^*_{\mu\nu}  \ . \label{eq:5.12}
\end{equation}
The universal coupling of matter to $g_{\mu\nu}$ means that
(non-gravitational) laboratory rods and clocks measure this metric. [It
would take a purely gravitational clock, e.g.  that defined by the
orbital motion of two black holes, to measure the metric $g^*_{\mu\nu}$].
The action (\ref{eq:5.10}) contains one dimensionful constant $G_*$
(``bare" Newtonian constant) and several free functions:  the $n(n-1)/2$
arbitrary functions $\gamma_{ab} (\varphi)$ entering a general
$(\sigma$-model) metric in the $n$-dimensional space of scalar fields
$(d\sigma^2 = \gamma_{ab} (\varphi^c) d\varphi^a d\varphi^b)$ and the
two functions $A(\varphi^a)$ and $B(\varphi^a)$ which give the coupling
of the scalars to the matter, and the self-couplings (potential) of the
scalars respectively. The original theory of Jordan-Fierz-Brans-Dicke
\cite{JFBD} has only one scalar field and one free parameter $,\alpha$.
This theory is defined by the choices $A(\varphi)=\exp (\alpha\varphi)$,
$B(\varphi)=0$, $d\sigma^2 =(d\varphi)^2$. The coupling parameter
$\alpha =\partial \ln A/\partial\varphi$ (which is a constant in this
theory) is related to the often quoted parameter $\omega$ through
$\alpha^2 =(2\omega +3)^{-1}$.

The gravitational field equations corresponding to the action
(\ref{eq:5.10}) read
\begin{eqnarray}
&& R^*_{\mu\nu} = 2\gamma_{ab} (\varphi) \partial_\mu \varphi^a
\partial_\nu \varphi^b + 2 B (\varphi) g^*_{\mu\nu}
+ 2q_* \left( T^*_{\mu\nu} - {1\over 2} T^* g^*_{\mu\nu} \right)
\ , \label{eq:5.12a} \\
&& \hbox{\rlap{$\sqcap$}$\sqcup$}_{g_*} \varphi^a + g^{\mu\nu}_*
\gamma^a_{bc} (\varphi) \partial_\mu \varphi^b \partial_\nu \varphi^c
- \gamma^{ab} (\varphi) {\partial B\over \partial \varphi^b} =
- q_* \alpha^a (\varphi) T_* \ . \label{eq:5.12b}
\end{eqnarray}
In eqs.~(\ref{eq:5.12a}), (\ref{eq:5.12b}) we have used the notation
\begin{eqnarray}
q_* &\equiv & 4 \pi G_*/c^4 \ , \label{eq:5.13a} \\
T^{\mu\nu}_* &\equiv & {2c\over \sqrt{g_*}}\,
{\delta S_m [\psi_m, A^2 g^*_{\mu\nu}] \over \delta g^*_{\mu\nu} }
\ , \label{eq:5.13b} \\
\alpha_a (\varphi) &\equiv & {\partial\, \ln A(\varphi) \over
\partial \varphi^a}\,\equiv \, A^{-1}\,
{\partial A\over \partial\varphi^a}\ . \label{eq:5.13c}
\end{eqnarray}
Moreover, $\hbox{\rlap{$\sqcap$}$\sqcup$}_{g_*} \equiv g^{\mu\nu}_*
\nabla^*_\mu \nabla^*_\nu$ denotes the $g_*$-covariant Laplacian,
$\gamma^{ab}$ the inverse of $\gamma_{ab}$, $\gamma^a_{bc}$ the
Christoffel coefficients of $\gamma_{ab}$, and the various indices are
moved by their corresponding metric: $T^*_{\mu\nu} \equiv
g^*_{\mu\alpha} g^*_{\nu\beta} T^{\alpha\beta}_*$, $\alpha^a \equiv
\gamma^{ab} \alpha_b$, etc.

Note that the ``Einstein-conformal-frame" stress-energy tensor
(\ref{eq:5.13b}) is related through
\begin{equation}
T^{\mu\nu}_* = A^6 T^{\mu\nu}\ , \ \sqrt{g_*} T^\nu_{*\mu} =
\sqrt g T^\nu_\mu \label{eq:5.14}
\end{equation}
(in which $T^\nu_\mu \equiv g_{\mu\alpha} T^{\alpha\nu})$ to the
physical (``Jordan-Fierz-frame") stress-energy tensor
\begin{equation}
T^{\mu\nu} \equiv {2c\over \sqrt g}\, {\delta S_m [\psi_m, g_{\mu\nu}]
\over \delta g_{\mu\nu} } \ . \label{eq:5.15}
\end{equation}
The latter tensor satisfies
\begin{equation}
\nabla_\nu T^{\mu\nu} = 0 \ , \label{eq:5.16}
\end{equation}
with respect to the $g$-covariant derivative $\nabla_\mu$, while it is
only the sum of $T^{\mu\nu}_*$ and of the stress-energy tensor of the
scalar fields which is $g_*$-covariantly conserved.

One sees from eq.~(\ref{eq:5.12a}) that the scalar potential
$B(\varphi)$ introduces an effective cosmological constant in the
tensorial field equations, $\Lambda = 2 <B(\varphi)>.$ There are very
tight constraints on the value of $\Lambda$ in ordinary units. One
should therefore restrict oneself to considering potentials $B(\varphi)$
that tend to dynamically drive the scalar fields toward values
$\varphi^a_0$ at which $B(\varphi^a_0)=0$. From eq.~(\ref{eq:5.12b}) we
see that this means, if $\gamma_{ab}$ is positive definite, that
$B(\varphi)$ should have a zero minimum value at $\varphi^a =
\varphi^a_0$. Setting by convention $\varphi^a_0=0$, we can then easily
write down the linearized approximation of the field equations:
\begin{eqnarray}
- 2 [R^*_{\mu\nu}]^{\rm linearized} \equiv &&
\hbox{\rlap{$\sqcap$}$\sqcup$}_* h^*_{\mu\nu} + \partial^*_{\mu\nu}
h^{*\alpha}_{\alpha} - \partial^*_{\alpha\mu} h^{*\alpha}_\nu
- \partial^*_{\alpha\nu} h^{*\alpha}_\mu \nonumber \\
&& = - 4q_* \left( T^*_{\mu\nu} - {1\over 2}\, T^*f^*_{\mu\nu}\right)
\ , \label{eq:5.17a} \\
(\hbox{\rlap{$\sqcap$}$\sqcup$}_* - m^{*2}_a) \varphi^a = && -q_*
\alpha^a (0) T_* \ . \label{eq:5.17b}
\end{eqnarray}
Here we have expanded the Einstein metric as $g^*_{\mu\nu} = f^*_{\mu\nu}
+ h^*_{\mu\nu}$ where $f^*_{\mu\nu}$ is a flat metric (which takes the
usual Minkowskian form when using some Einstein-frame coordinates
$x^\mu_*$); $\partial^*_\mu$ denotes $\partial/\partial x^\mu_*$,
$\hbox{\rlap{$\sqcap$}$\sqcup$}_* \equiv f^{\mu\nu}_* \partial^*_{\mu\nu}$,
and we used field coordinates $\varphi^a$ that diagonalize the scalar
mass matrix, i.e. the second-order gradients of $B(\varphi)$ around
zero, $\gamma^{bc} [\partial^2 B/\partial \varphi_a \partial\varphi_c]_0
= m^{*2}_a \delta^b_a$.

Inserting the solutions of eqs.~(\ref{eq:5.17a}) and (\ref{eq:5.17b})
into the action (\ref{eq:5.6}) giving the interaction between the matter
and the gravitational fields $h^*_{\mu\nu}$ and $\varphi^a$, namely
\begin{equation}
S^{\rm linearized}_{\rm int} = {1\over 2} \int {d^4x_*\over c}
\, (h^*_{\mu\nu} T^{\mu\nu}_* + 2 \alpha_a (0) \varphi^a T_*)\ ,
\label{eq:5.18}
\end{equation}
gives [using the harmonic gauge $\partial_*^\nu (h^*_{\mu\nu} - {1\over
2} h^{*\alpha}_\alpha f_{\mu\nu}) = 0]$
\begin{eqnarray}
S^{\rm linearized}_{\rm int} = -{4\pi G_*\over c^4} 
\int{d^4x_* \over c}  &&\left[
T^{\mu\nu}_{*{\rm loc}} \hbox{\rlap{$\sqcap$}$\sqcup$}_*^{-1}
( 2T^{*{\rm ext}}_{\mu\nu} - T^{\rm ext}_* f^*_{\mu\nu} ) \right.
\nonumber \\
&& \left.\qquad +\sum^{n}_{a=1} \alpha_a (0) \alpha^a (0) T_*^{\rm loc}
(\hbox{\rlap{$\sqcap$}$\sqcup$}_* - m^{*2}_a)^{-1} T_*^{\rm ext}
\right]\ ,        \label{eq:5.19}
\end{eqnarray}
where $T^{*{\rm loc}}_{\mu\nu}$ is the energy distribution of a local
system which is gravitationally interacting with the external energy
distribution $T^{*{\rm ext}}_{\mu\nu}$. Eq.~(\ref{eq:5.19}) shows
clearly that the metric $g^*_{\mu\nu}$ mediates a usual, Einstein-type
massless spin-2 interaction, while each scalar field mediates a,
possibly massive, spin-0 interaction.

The Newtonian limit of eq.~(\ref{eq:5.19}) consists in neglecting all
velocity dependent terms, which amounts to neglecting the components
$T^{0i}_*$ and $T^{ij}_*$ with respect to the time-time components
$T^{00}_*$. [Indeed, for ordinary materials $T^{00}_* \sim \rho c^2$,
$|T^{0i}_*| \sim \rho cv$, $|T^{ij}_*| \sim \rho v^2$ where $v$ is
some (internal or orbital) velocity]. This yields the following
interaction Lagrangian between two (point-like) bodies
\begin{equation}
L^{\rm Newtonian}_{\rm int} = G_* A(0)^2\, {M_1 M_2\over r_{12}}
\left[ 1 + \sum^n_{a=1} \alpha_a (0) \alpha^a(0)\, e^{-m_a r_{12}}
\right] \ , \label{eq:5.20}
\end{equation}
where the factor $A(0)^2$ comes from having rescaled both the stress-energy
tensor and the coordinates when passing from the Einstein frame
$x_*^\mu$ to the physical frame $x^\mu$, such that $ds^2 = g_{\mu\nu}
dx^\mu dx^\nu = A^2(\varphi) g^*_{\mu\nu} dx^\mu_* dx^\nu_*$ tends to
the usual Minkowski metric $f_{\mu\nu}dx^\mu dx^\nu$ at infinity.  The
scaling transformations are $x^\mu = A(0)x^\mu_*$ and eq.~(\ref{eq:5.14})
must be modified by taking into account the coordinate change $x^\mu_* \to
x^\mu$.  [As it stands eq.~(\ref{eq:5.14}) assumes the use of the same
coordinates in the Einstein and Jordan-Fierz frames].  The quantities
appearing in the final eq.~(\ref{eq:5.20}) are all expressed in physical
units [e.g.  $M = \int d^3 x T^{00}/c^2$ using physical coordinates
$x^\mu$ and the stress tensor (\ref{eq:5.15})].

The most evident experimental consequence of the result (\ref{eq:5.20})
[besides its pure dependence on the total mass-energies (equivalence
principle)] is the possible presence of Yukawa-type modifications of
the usual $1/r$ potential. Many experiments have set tight constraints
on such modifications.  Here is a sample of some recent results
\cite{A91,M93}, assuming the presence of only one Yukawa term (with
range $\lambda \equiv 1/m$)
\begin{eqnarray}
|\alpha (0)|^2 &\lessim& 10^{-4} \quad{\rm if}\quad
\lambda\simeq 1\ {\rm cm} \ ,  \nonumber \\
|\alpha (0)|^2 &\lessim& 5\times 10^{-4} \quad{\rm if}\quad
\lambda\simeq 1\ {\rm m} \ ,  \nonumber \\
|\alpha (0)|^2 &\lessim&  10^{-3} \quad{\rm if}\quad
10\ {\rm m} \leq \lambda \leq  10\ {\rm km} \ ,  \nonumber \\
|\alpha (0)|^2 &\lessim&  10^{-8} \quad{\rm if}\quad
10^4\ {\rm km} \leq \lambda \leq  10^5\ {\rm km} \ . \label{eq:5.21}
\end{eqnarray}
[Beware that the coefficient of the Yukawa term, here denoted $[\alpha
(0)]^2$, because it appeared as the square of the coupling constant
$\alpha (0)$ of the scalar field, is usually denoted $\alpha$]. For the
same reasons that we evoked above in the case of composition-dependent
interactions, it seems desirable to continue performing experiments,
both in the 10m-10km window, where the limits are not very
stringent and in the $\lambda <1$ mm window which is very poorly constrained
\cite{MP88}.

In the rest of these lectures, we shall concentrate on the case where
there are only long-range scalar fields ($m_a =0$). In that case
eq.~(\ref{eq:5.20}) predicts a $1/r$ potential between two masses
with an effective Newtonian constant given by
\begin{equation}
G= G(\varphi^a_0) = G_* [A(\varphi_0)]^2 [1 + \alpha^2 (\varphi_0)]
\ , \label{eq:5.22}
\end{equation}
where
\begin{equation}
\alpha^2 (\varphi) \equiv \gamma^{ab} (\varphi) \alpha_a (\varphi)
\alpha_b (\varphi) \equiv \gamma^{ab} \,
{\partial \ln A\over \partial \varphi^a}\, {\partial \ln A\over
\partial \varphi^b} \label{eq:5.23}
\end{equation}
denotes the fractional contribution of all the scalar fields to the
$1/r$ interaction. We have made explicit in eq.~(\ref{eq:5.22}) the
dependence of the effective Newtonian constant (as it can be measured by
a local Cavendish experiment) upon the background value (or VEV) of the
scalar fields. Indeed, as we are now considering the case of massless
scalars there is no need to assume any non-zero potential function $B
(\varphi)$ for the scalars. In that case, there is no longer a force term
$\propto -\partial B/\partial \varphi^a$ in eq.~(\ref{eq:5.12b}) driving
the scalars to a particular VEV $\varphi^a_0$.  On the contrary, the
long-range coupling of the scalar fields to the universe at large, now
exhibited by eq.~(\ref{eq:5.12b}), makes us expect that localized
gravitational systems will be embedded in a cosmologically evolving
background:  $\varphi^a_0 (t)$.  Therefore we expect from
eq.~(\ref{eq:5.22}) that the locally measured gravitational constant
will evolve on a Hubble time scale
\begin{equation}
(dG /dt) / G \sim H_0 \ . \label{eq:5.24}
\end{equation}
Various types of observational data
(including binary pulsar data \cite{42a} )
can be used to look for a possible
time-variation of the Newtonian coupling constant. Let us only quote
here the result obtained by a recent re-analysis of the Viking data
\cite{S90} (radar ranging between the Earth and Mars)
\begin{equation}
\dot G/G = (- 0.2 \pm 1.0) \times 10^{-11} {\rm yr}^{-1}\ .\label{eq:5.25}
\end{equation}

As $H_0 = h_{75} \times 75\ {\rm km/s}\, {\rm Mpc} = h_{75} \times 7.67
\times 10^{-11} {\rm yr}^{-1}$ with $h_{75} =1\pm 0.33$, we see by
comparing (\ref{eq:5.24}) and (\ref{eq:5.25}) that the present
observational results are not putting a very strong constraint on the
possible existence of a long-range coupling to the universe at large.
In fact the analysis of the post-Newtonian effects in the dynamics of
the solar system (see below) are putting much more severe constraints on
the existence of extra long-range fields than the present $\dot G$
observations.  [This is the case if one assumes a universal metric
coupling.  If, on the other hand, the matter driving the cosmological
expansion is a new type of (dark) matter which couples differently to a
postulated long-range scalar field the $\dot G$ observations may provide
a significant constraint on the scalar coupling of this dark matter
\cite{DGG}.]

To conclude this section devoted to the Newtonian limit let us recall
the shameful fact that Newton's  gravitational constant is one of the least
precisely measured fundamental constant of physics. By contrast
to $\hbar$, $\alpha = e^2 /\hbar c$, the particle masses,\dots which
are known with a part in a million precision (or better), $G$ is only
known with a precision $\sim 1.3 \times 10^{-4}$ \cite{CT87}:
\begin{equation}
G^{\rm obs} = [6.67259 \pm 0.00085] \times 10^{-8} {\rm cm}^3
{\rm g}^{-1} {\rm s}^{-2} \ . \label{eq:5.26}
\end{equation}
This lack of precision could become very annoying if ever theoretical
physics allows us, one day, to predict the value of $G$ in terms of
other physical constants.  Landau \cite{L55} entertained this hope long
ago and conjectured that the very small dimensionless quantity
$Gm^2/\hbar c \sim 10^{-40}$, where $m$ is a typical particle mass, might
be connected with the fine-structure constant $\alpha = [137.0359895
(61)]^{-1}$ by a formula of the type $A \exp (-B/\alpha)$, where $A$ and
$B$ are numbers of order unity.  Recently, 't~Hooft \cite{tH}
resurrected this idea in the context of instanton physics, where such
exponentially small factors appear naturally.  He went further in
suggesting (for fun) specific values for $A$ and $B$ in the case where
$m$ is the electron mass.  Actually, the final formula he proposed is in
significant disagreement with the observed value (\ref{eq:5.26}).
However, keeping his (instanton-motivated) value for $B$, namely
$B = \pi/4$, but taking for $A$ the value $(7\pi)^2 /5$ one can (still
for fun) define a simple-looking ``theoretical" value for $G$ by $G^{\rm
theory} m^2_e/\hbar c \equiv (7\pi)^2/5\, \exp (-\pi/ 4\alpha).$ Using
the central values of the 1986 adjustment of the fundamental physical
constants \cite{CT87}, this formula ``predicts" $G^{\rm theory} =
[6.6723458\cdots] \times 10^{-8} {\rm cm}^3{\rm g}^{-1} {\rm s}^{-2}$,
which is in good agreement with the observed value (\ref{eq:5.26}):
$G^{\rm obs} /G^{\rm theory} = 1.00004 \pm 0.00013$~!  Let this exercise
serve as a reminder of the potential importance of improving the
precision of the measurement of $G$.

\subsection{The post-Newtonian limit of tensor-multi-scalar theories and
its experimental tests.} \label{sec:5.3}

The term ``post-Newtonian" refers to the terms in the  Lagrangian
describing  the motion of gravitationally interacting bodies which
contain a factor $1/c^2$ with respect to the ``Newtonian" terms
(\ref{eq:5.20}). There are two types of post-Newtonian terms: those
which are smaller than (\ref{eq:5.20}) by a factor $(v/c)^2$
[``velocity-dependent terms"], and those which are smaller by a factor
$GM/rc^2$ [``non-linear terms"].

The velocity-dependent terms (also called ``gravitomagnetic" terms)
can be directly deduced from the linearized-order action (\ref{eq:5.19})
by inserting the point-mass approximation of $T^{\mu\nu} = \int m u^\mu
u^\nu \delta (x-z(s)) ds$ (after the needed scaling transformations).
The latter equation shows very clearly that the exchange of massless
scalar fields introduces a different velocity dependence $\propto
\int\!\!\int ds_1 ds_2 m_1 m_2 G(z_1 - z_2)$ than the one due
to the exchange of a massless spin-2 field $\propto \int\!\!\int ds_1
ds_2 m_1 m_2 u_{1\mu} u_{1\nu} (2u_2^\mu u^\nu_2 + f^{\mu\nu}) G(z_1-z_2)$
[here $G(x)$ denotes the Green function of the flat-space d'Alembertian,
and $u^\mu_1$, $u^\mu_2$ the four-velocities of the two considered mass
points].  Similarly to what happened in eqs.~(\ref{eq:5.20}) and
(\ref{eq:5.22}) one sees from eq.~(\ref{eq:5.19}) that (in the massless
case) the factor $\alpha^2$ of eq.~(\ref{eq:5.23}) will weigh the
contribution of the scalars to the velocity-dependent terms [see Sec.~3
of Ref.~\cite{DEF1} for details].

Let us now turn our attention to the non-linear post-Newtonian terms
$\propto GM/c^2r$. There are two types of such terms. The first type
can be easily understood from our previous results.  Indeed,
eq.~(\ref{eq:5.22}) showed that the value of the gravitational coupling
constant measured in a local Cavendish experiment depends upon the
ambient values of the externally generated scalar fields at the location
where the experiment is performed.  [The calculation behind
eq.~(\ref{eq:5.22}) considered a gravitating system put in a constant
scalar background $\varphi^a_0$ (of cosmological origin).  Because of
the long range of the scalars, the scalar background experienced by one
body member of an $N$-body system is obtained by adding the effects of
the $N-1$ other bodies onto the cosmological background].  Therefore the
effective gravitational constant ruling the self-gravity of a particular
body (say a planet) will be space dependent:  $G({\bf x}) = G(\varphi
({\bf x})),$ where $G(\varphi)$ is given by eq.~(\ref{eq:5.22}).  Now,
the total mass-energy of a self-gravitating body depends upon $G$
because of the gravitational binding energy, say $E^{\rm grav} \equiv G
\partial (m^{\rm tot} c^2) /\partial G \not= 0.$ The space-dependence of
$G({\bf x})$ induces a space-dependence of the mass $m$.  As was
discussed in Sec.~\ref{sec:4.2.1} above, [eqs.~(\ref{eq:4.7}) and
(\ref{eq:4.8})], this causes a supplementary term in the acceleration of
the body, namely
\begin{eqnarray}
(\delta {\bf a})^{\rm self-gravity} &=& -\mbox{\boldmath{$\nabla$}}\ln m
= -{\partial \ln m\over \partial \ln G}\,
\mbox{\boldmath{$\nabla$}}\ln G \nonumber \\
&=& - {E^{\rm grav}\over mc^2}\, {\partial \ln G\over \partial\varphi^a}
\mbox{\boldmath{$\nabla$}} \varphi^a \ . \label{eq:5.29}
\end{eqnarray}
Such a term is absent in pure general relativity where the gravitational
influence of the external universe can be locally effaced by introducing
Fermi-Cartan coordinates, eq.~(\ref{eq:4.3}). [See Ref.~\cite{D87} for
a general discussion of the ``effacement" properties present in general
relativity, and for references]. The presence in tensor-scalar theories
of an anomalous contribution to the gravitational acceleration of a body
proportional to $E^{\rm grav}/mc^2$ was discovered by Nordtvedt
\cite{N68a}. [The possibility of such an effect was first noticed, via
the reasoning behind eq.~(\ref{eq:5.29}), by Dicke \cite{D64b}].

The second type of non-linear terms are the genuine 3-body interaction
terms in the action for gravitating bodies. To obtain them one needs
to go beyond the linearized theory written down in eqs. (\ref{eq:5.17a}),
(\ref{eq:5.17b}) above, and study the quadratically non-linear terms in the field
equations, i.e.  the cubic terms in the field action (\ref{eq:5.10})
[Fortunately, it suffices to study these terms in the slow-motion limit].  An
elegant way of dealing with these quadratic nonlinearities has been recently
found both in general relativity \cite{DSX} and in tensor-multi-scalar
theories \cite{DEF1}.  Let us quote the final result for the Lagrangian
describing, within the first post-Newtonian approximation, the
gravitational dynamics of $N$ (self-gravitating) bodies [with positions
${\bf z}_A(t)$ and velocities ${\bf v}_A (t)$;  $A=1-N$]
\begin{equation}
L^{\rm N-body} ({\bf z}_A, {\bf v}_A) = \sum_A L^{(1)}_A +
{1\over 2}\, \sum_{A\not= B} L^{(2)}_{AB} + {1\over 2}
\sum_{B\not= A\not= C} L^{(3)A}_{BC} + O(c^{-4}) \ , \label{eq:5.30}
\end{equation}
where
\begin{eqnarray}
L^{(1)}_A  && =  -m_A c^2 \sqrt{1-{\bf v}^2_A /c^2} \nonumber \\
&& = -m_A c^2 +{1\over 2} m_A {\bf v}^2_A + {1\over 8c^2} m_A
({\bf v}^2_A)^2 + O \left( {1\over c^{4}} \right)\ , \label{eq:5.31a}\\
L^{(2)}_{AB} && = {Gm_Am_B\over r_{AB}} \left[ 1 + (4\beta - \gamma -3)
\left( {E^{\rm grav}_A\over m_Ac^2}+{E^{\rm grav}_B\over m_Bc^2}\right)
+ {\gamma -1\over c^2} ({\bf v}_A -{\bf v}_B)^2 \right. \nonumber \\
&& \left. \qquad \qquad + {3\over 2c^2} ({\bf v}^2_A + {\bf v}^2_B)
- {7\over 2c^2} ({\bf v}_A \cdotp {\bf v}_B) - {1\over 2c^2}
({\bf n}_{AB} \cdotp {\bf v}_A) ({\bf n}_{AB} \cdotp {\bf v}_B) \right]
\ , \label{eq:5.31b} \\
L^{(3)A}_{BC} && = -(1+2(\beta -1)){G^2m_Am_Bm_C\over c^2r_{AB}r_{AC}}
\ , \label{eq:5.31c}
\end{eqnarray}
with $r_{AB} = |{\bf z}_A - {\bf z}_B|$ and ${\bf n}_{AB} = ({\bf z}_A
-{\bf z}_B) /r_{AB}$.  The physical metric corresponding to the
post-Newtonian level of accuracy can be written as [using the short-hand
notation $O(n) \equiv O(c^{-n})]$
\begin{eqnarray}
g_{00} &=& -\exp\left[ -{2\over c^2}V+2(\beta -1){V^2\over c^4} \right]
+ O(6) \ , \label{eq:5.32a} \\
g_{0i} &=& - {2\over c^3} (\gamma +1) V_i + O(5) \ , \label{eq:5.32b}\\
g_{ij} &=& +\exp \left[ +{2\over c^2}\gamma V \right]\delta_{ij}
+ O(4)\ ,   \label{eq:5.32c}
\end{eqnarray}
in terms of the following scalar and vector potentials
($\hbox{\rlap{$\sqcap$}$\sqcup$} \equiv f^{\mu\nu} \partial_{\mu\nu})$
\begin{eqnarray}
\hbox{\rlap{$\sqcap$}$\sqcup$} V &=& - 4\pi G \left[ 1
+ (3\gamma -2\beta -1) {V\over c^2} \right] {T^{00}+T^{ss} \over c^2}
\ , \label{eq:5.33a} \\
\hbox{\rlap{$\sqcap$}$\sqcup$} V_i &=& - 4\pi G \, {T^{0i}\over c} \ .
\label{eq:5.33b}
\end{eqnarray}
[One should keep in mind that the post-Newtonian limit is a combined
weak-field, slow-motion expansion, so that the error terms
$O(n) = O(c^{-n})$ in eqs.~(\ref{eq:5.32a})--(\ref{eq:5.32c}) contain
both velocity-dependent terms (or time derivatives), and higher-order
nonlinear terms].  Besides the (dimensionful) constant $G$,
eq.~(\ref{eq:5.22}), there enters only two (dimensionless) parameters in
the post-Newtonian limit of tensor-multi-scalar theories:  $\gamma$ and
$\beta$.  [They coincide with the parameters introduced by Eddington
long ago when considering the simpler model of test particles moving in
the field of one central, massive body \cite{E22}].  The post-Newtonian
limit of general relativity is obtained when $\gamma = 1 =\beta$.  [Note
the simplifications of the non linear structure that arise in this limit
where $4\beta -\gamma-3 =0=3\gamma -2\beta -1].$

The quantity $\gamma-1$ parametrizes the possible presence of
non-general-relativistic velocity-dependent terms (see
eq.~(\ref{eq:5.31b})). From our discussion above it is clear that
$\gamma -1$ must be proportional to $\alpha^2$, eq.~(\ref{eq:5.23}),
which measures the admixture of the scalars in the two-body interaction.
More precisely, one finds
\begin{equation}
\gamma -1 = - 2 { \alpha^2\over 1+\alpha^2} \ . \label{eq:5.34}
\end{equation}
The result (\ref{eq:5.34}) can be formally generalized to the case
where the gravitational interaction is mediated not only by (massless)
spin-2 and spin-0 fields, but also by (massless) spin-1 fields. [This
generalization is formal because, as we saw above, spin-1 fields cannot
couple exactly to the mass]. If $g_s$ denotes the coupling constant
of spin-s fields one finds \cite{DEF1}
\begin{equation}
G = \sum_s (-)^s g^2_s \ , \label{eq:5.35}
\end{equation}
\begin{equation}
\gamma +1 = {1\over 2} \,{\sum_s (-)^s g^2_s s^2\over\sum_s (-)^s g^2_s}
= {1\over 2} <s^2> \ . \label{eq:5.36}
\end{equation}
[We denote here $g^2_s$ what was denoted $g_s$ in Ref.~\cite{DEF1}].
Note the elegant interpretation of $\gamma +1$ as being half the average
squared spin of the mediating fields [the weights being defined by the
contributions of the fields to the 2-body interaction, including the sign
which is negative (repulsion) for $s=1$]. When $g_1=0$ and $g_0/g_2 =
\alpha^2$, eq.~(\ref{eq:5.36}) yields (\ref{eq:5.34}).

The quantity $\beta -1$ parametrizes the possible deviations from
general relativity in the non-linear terms. Its expression in a general
tensor-multi-scalar theory is
\begin{equation}
\beta -1 = {1\over 2}\, {\alpha^a \beta_{ab} \alpha^b \over
[1 +\alpha^2]^2} \ , \label{eq:5.37}
\end{equation}
where
\begin{equation}
\beta_{ab} \equiv D_a D_b \ln A = D_a \alpha_b = \partial_a \alpha_b
- \gamma^c_{ba} \alpha_c    \label{eq:5.38}
\end{equation}
is the second covariant derivative (with respect to the $\sigma$-model
metric $\gamma_{ab}$) of the logarithm of the coupling function
$A(\varphi^a)$.

Many observations in the solar system have been used to study the
post-Newtonian effects present in eqs.~(\ref{eq:5.30})-(\ref{eq:5.33b}). [A
famous example is the secular advance of the perihelion of Mercury, already
discussed in Sec.~\ref{sec:2}]. At present, two sorts of experiments stand out as
giving the tightest constraints on $\gamma$ and $\beta$. [See
\cite{W81,S90} for a discussion of the other tests of post-Newtonian
gravity]. Time-delay measurements \cite{S64} based on the Viking ranging data
to Mars \cite{R79}, and Very-Long-Baseline-Interferometry measurements of the
deflection of radio waves by the Sun \cite{RCD91}, \cite{L+95}, have allowed one
to measure $\gamma$, (nearly) independently from $\beta$. In the former case,
this is done by considering the time-of-flight of an electromagnetic (radar)
signal sent from the Earth, actively reflected on a Viking lander on the surface
of Mars, and received back on Earth.  Writing from eqs.~(\ref{eq:5.32a},
\ref{eq:5.32b}, \ref{eq:5.32c}) the curved-space equation for the light cone,
$0=ds^2=g_{\mu\nu} dx^\mu dx^\nu$, one finds that the coordinate time of flight
$(x^0=ct)$ is given with sufficient accuracy by 
\begin{equation} 
c\Delta t \simeq \int \left[ 1+
(1+\gamma) {V\over c^2} \right] |d{\bf x}|
\ , \label{eq:5.39}
\end{equation}
where $V \simeq GM_\odot /r$. Using an accurate ephemeris for predicting
the coordinate positions of the Earth and Mars at each coordinate time
$t$, and the transformation between $\Delta t$ and the proper time
measured by Earth clocks, one can measure the coefficient $1+\gamma$ of
the relativistic (or ``Shapiro" \cite{S64}) time delay $\int V |d{\bf
x}|/c^3$ appearing in eq.~(\ref{eq:5.39}) by analyzing the data where
the electromagnetic signals pass near the Sun.  [The specific time
signature of the ``Shapiro" time delay allows one to separate it from
the many other effects present in the leading ``Roemer" time of flight
$\int |d{\bf x}|/c$].  The final result of this Viking time-delay
experiment is \cite{R79}
\begin{equation}
\gamma = 1.000 \pm 0.002 \ . \label{eq:5.40}
\end{equation}

The same limit was found in the deflection experiment \cite{RCD91}, while a
very recent deflection measurement obtained a slightly better limit: $\gamma =
0.9996 \pm 0.0017$ \cite{L+95}.

The second high-precision test of post-Newtonian gravity comes from an
analysis of the laser ranging data to the Moon \cite{M77}.  In July
1969, the Apollo~11 mission, besides its spectacular aspect of having
landed the first men on the Moon, left a panel of corner-cube reflectors
on the surface of the Moon.  Since August~1969 one has accumulated
regular measurements of the round trip travel times of laser pulses sent
from several stations on the Earth [two at present;  CERGA, France and
McDonald, USA] and bounced off an array of 4 lunar reflectors (the first
Apollo~11 reflector has been completed by two other US-made reflectors
--~Apollo~14 and 15~-- and by two French-made reflectors deposited by
the Russian Lunakhod 1 and 2 missions.  Alas the Lunakhod 1 reflector
never sent back any echo).  Because the Earth and the Moon have
non-negligible gravitational binding energies $[(E^{\rm grav}/mc^2)_{\rm
Earth} \simeq -4.6 \times 10^{-10}$, $(E^{\rm grav}/mc^2)_{\rm Moon}
\simeq -0.2 \times 10^{-10}]$, eq.~(\ref{eq:5.29}) shows that they could
fall with a different acceleration towards the Sun.  Computing $\partial
\ln G/\partial \varphi^a$ from eq.~(\ref{eq:5.22}), and ${\bf \nabla}
\varphi^a$ from eq.~(\ref{eq:5.17b}) one finds
\begin{equation}
(\delta {\bf a})^{\rm self-gravity} = (4\beta -\gamma -3)\,
{E^{\rm grav}\over mc^2}\, \mbox{\boldmath{$\nabla$}}V \ .\label{eq:5.41}
\end{equation}
One recognizes here the effect of the second term on the
right-hand-side of eq.~(\ref{eq:5.31b}). Eq.~(\ref{eq:5.41}) means
that the combination $(4\beta -\gamma -3)$ parametrizes the violation
of the universality of free fall happening for self-gravitating bodies
in theories that differ from general relativity (``violation of the
strong equivalence principle"). This effect was discovered by
Nordtvedt \cite{N68a}, who emphasized also that laser ranging to the
Moon offered an excellent way of looking for the presence of the term
(\ref{eq:5.41}) \cite{N68b}.
[Note  that we are working here under the assumption that there is no
violation of the ``weak'' equivalence principle associated with 
the different compositions of the Earth and the Moon].
Indeed, the differential acceleration of
the Earth-Moon system in the field of the Sun induces a polarization of
the Moon's orbit about the Earth. This consequence of a violation of the
equivalence principle was, in fact, first pointed out by Newton, see
section~6.6 of \cite{D87}, and first correctly worked out by Laplace
\cite{Lap}. For recent theoretical studies of this effect taking into account
the important mixing with solar tidal distortion see \cite{N95}, \cite{DVmoon}.
The most recent analyses of the experimental data yield \begin{equation}
 4\beta - \gamma -3 = -0.0005 \pm 0.0011  \label{eq:5.42}
\end{equation}
according to Ref.~\cite{D+94} and $4\beta -\gamma -3=-0.0007 \pm 0.0010$
according to Ref.~\cite{WND95}.

Combining (\ref{eq:5.40}) and (\ref{eq:5.42}) yields the following
value for $\beta$
\begin{equation}
\beta = 0.9998 \pm 0.0006 \ . \label{eq:5.43}
\end{equation}

To end this section, one should mention the fact that Nordtvedt and
Will \cite{N68a,W71,WN72} have introduced a more general ``parametrized
post-Newtonian" formalism containing, besides the two parameters
$\gamma$ and $\beta$, eight other dimensionless parameters, $\xi$,
$\alpha_1$, $\alpha_2$, $\alpha_3$, $\zeta_1$, $\zeta_2$, $\zeta_3$,
$\zeta_4$, associated with other, a priori conceivable, deviations from
general relativity.  In particular, the two parameters $\alpha_1$ and
$\alpha_2$ are associated with a possible gravitational violation of
local Lorentz invariance (existence of preferred frames).  The original
theoretical motivation for considering such preferred-frame parameters
was the idea that gravity could be mediated in part by a long-range
vector field (or by other tensor fields).  If that were the case one
would expect the Universe's global matter distribution to select a
preferred rest frame for the gravitational interaction.  However, we saw
above that there were theoretical difficulties in constructing
consistent field theories that are metric and contain vector or tensor
fields.  The situation is even worse for the other PPN parameters that
do not seem to come out of any decent field theory.  It is anyway a
meaningful phenomenological question to ask whether all the existing
data about solar-system gravity suffice to put significant constraints
on all the PPN parameters.  The answer is yes for most of them
\cite{W81}.  Generally speaking the extra PPN parameters are much more
tightly constrained than $\gamma -1$ and $\beta -1$ (with the exception
of $\alpha_1$ which is presently constrained only at the level
$|\alpha_1| < 1.7 \times 10^{-4}$ \cite{H84}, \cite{DEF2}, \cite{BCD95} and
which needs new data to be more tightly constrained \cite{DEF94a},
\cite{DValpha}).

\subsection{Theoretical conclusions about weak-field, metric gravity}
\label{sec:5.4}

A first conclusion is that general relativity is consistent with all
existing tests of weak-field gravity (at the Newtonian and post-Newtonian
levels).  We have argued above that the most natural (and probably the
only theoretically consistent) metric alternative to Einstein's pure
spin-2 theory, is a metric theory where gravity couples exactly to mass
and is mediated both by one massless spin-2 and one or several, massive
or massless, spin-0 fields.  Within this framework the fraction of the
gravitational interaction carried by all the scalar fields is constrained
at the $10^{-3}$ level or better.  Indeed, in the case of finite-range
scalars one had the constraint $\alpha^2 \equiv \alpha_a \alpha^a <
10^{-3}$ from Newtonian measurements for ranges between 10~m and 10~km,
eq.~(\ref{eq:5.21}) (and tighter constraints for other ranges, except
for ranges $<1$~mm), while in the case of infinite-range scalars (or
with ranges greater than the Earth-Sun distance) the limit (\ref{eq:5.40})
yields (when using eq.~(\ref{eq:5.34})) the same numerical level
\begin{equation}
\alpha^2 < 10^{-3} \label{eq:5.44}
\end{equation}
for the constraint on a possible admixture of spin-0 in the gravitational
interaction.

Note that, while the existing observational limit on the post-Newtonian
parameter $\gamma -1$ yields a rather strong constraint on the most
natural theoretical alternatives to general relativity, the
observational limit (\ref{eq:5.43}) on the other phenomenologically
independent post-Newtonian parameter $\beta -1$ gives only a very modest
supplementary constraint on these alternative theories.  Indeed,
in the (most favourable) case where $\alpha^2 = \gamma_{ab} \alpha^a \alpha^b$
is equal to $10^{-3}$ the limit (\ref{eq:5.43}) is only telling us that the
largest eigenvalue of the matrix $\beta^a_b =\gamma^{ac} \beta_{cb}$,
eq.~(\ref{eq:5.38}), must be smaller than about 1.2.

These conclusions raise several questions. The first one is to know
what are the prospects for probing with higher precision possible
deviations from general relativity in the Newtonian (tests of $1/r^2$
law) and post-Newtonian regimes. Concerning the Newtonian tests there
is certainly room for improvement, coming e.g. from the use of
superconducting gravity gradiometers that can directly test whether the
Laplacian of the gravitational potential vanishes or not \cite{Paik}.

Concerning post-Newtonian tests, the Stanford gyroscope experiment
\cite{GPB} (now called GPB, for Gravity Probe B) aims at measuring the
velocity-dependent gravitational effects (``gravitomagnetism") with a
precision corresponding to the $10^{-5}$ level for $\gamma$, and thereby
$\alpha^2$.  A similar precision on $\gamma$ will be reached by the successor
of HIPPARCOS, the cornerstone project of ESA named GAIA. Some dedicated
missions (such as the high-precision time-transfer project SORT proposed to ESA
by C. Veillet et al.) might go down to the $\gamma -1 \sim 10^{-7}$ level.
The prospects for improving post-Newtonian orbital tests by using 
artificial satellites are discussed in ref. \cite{63a}. An improvement 
in post-Newtonian measurements, below the
present level (\ref{eq:5.44}) is important in the light of a recent
study of tensor-scalar cosmological models \cite{DN92}. It is found that
tensor-scalar metric theories generically contain a natural attractor
mechanism tending to drive the world toward a minimum of the coupling
function $A(\varphi)$, i.e.  toward a state close to a pure general
relativistic one $(\alpha_{\min} =(\partial \ln A/\partial\varphi)_{\min}
=0)$, with the redshift at the beginning of the matter-dominated era
providing the measure for the present level of deviation from general
relativity.  The numerical estimates of Ref.~\cite{DN92} indicate values
of $1-\gamma \grtsim 2(\Omega /0.1)^{-3/2} \times 10^{-5}$ where $\Omega
=\rho^{\rm matter} /\rho^{\rm critical}$ is the ususal dimensionless
measure of the average mass density in the universe.

Another natural question is to know whether it will become possible in
the future to probe weak-field gravity at the next level in the
post-Newtonian expansion, i.e.  at the $(v/c)^4$ level in the Lagrangian
(\ref{eq:5.30}) [``Second post-Newtonian level"]. Some theoretical studies
\cite{BN88,B92,DEF4} have generalized the parametrized post-Newtonian
formalism to the second post-Newtonian (2PN) level.  In particular, the
study of tensor-multi-scalar theories shows that there appear two and
only two new parameters at the $1/c^4$ level, namely \cite{DEF4}
\begin{eqnarray}
\beta_2 \equiv \alpha^a \beta^b_a \beta^c_b \alpha_c \ ,
\label{eq:5.45a} \\
\beta'  \equiv \alpha^a \alpha^b \alpha^c D_a \beta_{bc} \ .
\label{eq:5.45b}
\end{eqnarray}
>From a phenomenological point of view the two free 2PN parameters
$\beta_2$ and $\beta'$ can be considered as new independent parameters
whose values must be obtained, along with the values of $\gamma$ and
$\beta$, by fitting the data to a complete second-post-Newtonian-accurate
model of the solar system.  If the best fit gives $(\gamma, \beta,
\beta_2, \beta') = (1,1,0,0)$ that will confirm the validity of general
relativity at a deeper level than is presently achieved. Ref.\cite{DEF4} has
shown that binary pulsar data (see below) already give strong constraints on
these 2PN parameters: $| \beta_2 | < 6 \times 10^{-3}$, $|\beta'| < 7 \times
10^{-2}$. However, from a theoretical point of view, the results
(\ref{eq:5.45a},\ref{eq:5.45b}) tell us that such a deeper test (2PN versus 1PN)
is not really probing new, independent theoretical possibilities (at least among
the most natural alternatives to Einstein's theory).  Indeed, if we assume that
all the scalar fields carry only positive energy, which means that the
$\sigma$-model metric $\gamma_{ab}$ is positive definite, then the limit
(\ref{eq:5.44}), namely $\gamma_{ab} \alpha^a \alpha^b < 10^{-3}$, is severely
constraining the magnitude of all the individual coupling constants $\alpha^a$. 
As we expect $\beta_{ab}$ and $D_a \beta_{bc}$ to be of order unity we are led
to conclude already from the 1PN experimental results that $\beta_2$ and
$\beta'$ must be numerically small compared to one.  In other words, the message
seems to be that it is theoretically more important to put more effort in
determining the ``old" 1PN parameters, especially $\gamma$, rather than in
trying to look for 2PN deviations from general relativity.  [Practically
speaking, this means fitting the data to a theoretical description of gravity
given by the full 2PN limit of general relativity plus the 1PN deviations
parametrized by $\gamma -1$ and $\beta -1$].

\section{Testing the strong and radiative gravitational field regimes}
\label{sec:6}

\subsection{Binary pulsars as laboratories for probing strong and
radiative gravitational fields.} \label{sec:6.1}

All tests of relativistic gravity discussed above have been performed
within the solar system, i.e. within conditions characterized by a
slowly changing and extremely weak gravitational field. For instance
the relativistic gravitational potential of the Sun at the Earth
distance $GM_\odot /c^2 r_{\odot\oplus} \approx 10^{-8}$, which is of
the same order of magnitude as the velocity-dependent effects $\propto
(v^{\rm orbital}/c)^2 \sim [10^{-4}]^2.$ The measured relativistic
effects are but small perturbations to Newtonian expectations (as is
well expressed by the terminology of ``post-Newtonian" regime). In other
words, all the solar system tests have probed the gravitational
interaction only in the combined limit of weak and quasi-stationary
gravitational fields. Therefore, even when the experimental accuracy
is high these tests have an important qualitative weakness: they say
a priori nothing about how the ``correct" theory of gravity might behave
when the gravitational field is very strong (such as near a neutron
star or a black hole) or very rapidly varying (as in radiative
phenomena).

Fortunately the discovery of binary pulsars in 1974 \cite{HT75} opened
up an entirely new testing ground for relativistic gravity, giving us
for the first time an experimental handle on the regime of strong and/or
radiative gravitational fields.  Pulsars in gravitationally bound binary
orbits provide nearly ideal laboratories for the testing of strong-field
gravity:  being neutron stars, they have surface gravitational potentials
$GM/c^2R \approx G(1.4M_\odot) /c^2 (10\ {\rm km}) \approx 0.2$;  they
move with mildly relativistic velocities $(v/c \approx 10^{-3})$ through
a repetitive cycle well suited to experimental averaging techniques;
and they emit periodic pulses of radio noise, detectable over
interstellar distances, in some cases as stable as the ticks of an
atomic clock. The many orders of magnitude separating the
self-gravitational fields of pulsars $(GM/c^2R \approx 0.2$) from that
of the Earth $(GM_\oplus /c^2 R_\oplus \sim 10^{-9})$ or even the Sun
$(GM_\odot /c^2R_\odot \sim 10^{-6})$, and their closeness to the black
hole limit $((GM/c^2R)_{\rm BH} =0.5)$, make it clear that they give us
access to strong-field gravity.  The modest increase in orbital velocity
$((v/c)_{\rm PSR} \approx 10^{-3}$ versus $(v/c)_\oplus \approx
10^{-4})$ does not indicate clearly why they can also give us a handle
on radiative gravitational phenomena.  This comes from the fact that the
corresponding orbital periods $P_{\rm PSR}$ are of order of a fraction
of a day instead of a year, and that a binary pulsar is made of two objects
with comparable masses $\sim 1.4M_\odot$, while in the solar system
the planets are much less massive than the sun. Taking these two facts
into account one calculates easily that the change in orbital longitude
over some given time span, due to gravitational radiation damping, is
greater in a binary pulsar than in the orbit of the planet Mercury
by a factor $\sim (M_\odot /M_{\rm Mercury}) \times (P_{\rm Mercury}/
P_{\rm PSR})^{11/3} \sim 10^{16}$.

A last, but not least, advantage of binary pulsar systems over the solar
system is their theoretical simplicity and purity. On the one hand,
the solar system is a very complex dynamical system with many degrees
of freedom (many more than can be modelled in full detail; in particular
the modelling of asteroids is limiting the precision of many solar-system
relativistic tests).  On the other hand, a binary pulsar has essentially
only 6 degrees of freedom (although the spin degrees of freedom must be
accounted for).  In many respects, a binary pulsar is the hydrogen atom
of relativistic gravity, and like its electromagnetic analog it has
allowed one to investigate fine and hyper-fine levels of structure of
the gravitational interaction (including a classical surrogate of the
Lamb shift, i.e.  radiative effects in the orbital motion).

After the discovery of Hulse and Taylor \cite{HT75} many authors
realized the potentialities of binary pulsars for probing strong
and/or radiative gravitational fields. In the following, we summarize
the comprehensive approach of Damour and Taylor \cite{DT92} to which
we refer for details and references to earlier work. For reviews of the
use of pulsars as physics laboratories see the special issue of the
Philosophical Transactions of the Royal Society celebrating the 25th
anniversary of the discovery of pulsars \cite{Hewish}. See also the Nobel
lectures of Hulse and Taylor \cite{Nobel}.

\subsection{Phenomenological analysis of binary pulsar data
(``Parametrized Post-Keplerian Formalism")} \label{sec:6.2}

Binary pulsar data consist in the recording of the time of arrival,
shape and polarization of successive electromagnetic pulses emitted
by a pulsar member of a binary system. The binary nature of the system
implies that these data contain a wealth of information about
gravitational physics. Let us first consider the ``timing" data
(recording of the times of arrival of the centers of the pulses). The
intrinsic pulse mechanism is believed to be due to the spinning motion
of a neutron star. Like a rotating beacon atop a lighthouse, the
rotation of the magnetosphere structure of a neutron star sweeps a radar
beam across the sky. If this beam passes over the solar system, one
observes from the Earth a radio pulse for each turn of the pulsar. The
time of arrival on Earth of each pulse must be corrected for the Earth
motion around the Sun and for the dispersion due to the propagation
of the electromagnetic wave in the interstellar plasma. After having
done these corrections, the sequence of times of arrival gives us a
direct handle on the orbital motion of the pulsar. If the timing
precision is high (say, 1 to 10 microseconds) one can study in detail
many aspects of the relativistic two-body problem. More precisely it
has been shown that all the independent relativistic timing effects
bigger than or equal to $(v^{\rm orbital}/c)^2 P_b$ (where $P_b$ denotes
the binary period) can be described by a simple mathematical formula
common to a wide class of relativistic theories of gravity (the class of
boost-invariant metric theories, i.e. the tensor-multi-scalar ones).
This ``timing" formula \cite{DD} predicts that the time of arrival
(corrected for Earth motion and dispersion) of the $N$th pulse (where
$N$ is an integer) reads
\begin{equation}
t_N - t_0 = F [T, \{p^{\rm K}\}\, ;\, \{p^{\rm PK}\}\, ;\,
\{q^{\rm PK}\}]\ ,   \label{eq:6.1}
\end{equation}
where $T$ is the pulsar proper time (corrected for aberration)
corresponding to the $N$th turn, i.e.
\begin{equation}
N/2\pi = \nu_p T + {1\over 2}\, \dot\nu_p T^2 + {1\over 6}\, \ddot\nu_p
T^3 \label{eq:6.2}
\end{equation}
(where $\nu_p \equiv 1/P_p$ is the pulsar frequency), where
\begin{equation}
\{ p^{\rm K} \} = \{ P_b, T_0, e_0, \omega_0, x_0 \}
\label{eq:6.3}
\end{equation}
is the set of ``Keplerian" parameters,
\begin{equation}
\{p^{\rm PK}\} =
\{k, \gamma , \dot P_b, r, s, \delta_\theta, \dot e, \dot x \}
\label{eq:6.4}
\end{equation}
the set of {\it separately measurable} ``post-Keplerian" parameters
[among which $\gamma$ denotes a dimensionful time-dilation parameter to
be distinguished from the post-Newtonian parameter denoted by the same
letter],  and
\begin{equation}
\{ q^{\rm PK} \} =    \{ \delta_r, A, B, D \}    \label{eq:6.5}
\end{equation}
the set of {\it not separately measurable} ``post-Keplerian" parameters.
The right hand side of eq.~(\ref{eq:6.1}) is given by
\begin{eqnarray}
F(T) =&& D^{-1}
[ T + \Delta_R(T) + \Delta_E(T) + \Delta_S(T) + \Delta_A(T) ] \ ,
\label{eq:6.6a} \\
\Delta_R =&& x \sin \omega
[ \cos u - e (1 + \delta_r) ]
+ x [1 - e^2 (1+\delta_\theta)^2 ]^{1/2} \cos \omega \, \sin u \ ,
\label{eq:6.6b} \\
\Delta_E =&& \gamma \sin u \ , \label{eq:6.6c} \\
\Delta_S =&& - 2 r \, \ln
\{ 1 - e \cos u - s [\sin \omega (\cos u -e)
+ (1-e^2)^{1/2} \cos \omega \, \sin u] \} \ , \label{eq:6.6d} \\
\Delta_A =&& A \{ \sin [\omega + A_e (u)] + e \sin \omega \}
+ B \{ \cos [\omega + A_e (u)] + e \cos \omega \} \ , \label{eq:6.6e}
\end{eqnarray}
where
\begin{eqnarray}
x =&& x_0 + \dot x (T-T_0) \ , \label{eq:6.7a} \\
e =&& e_0 + \dot e (T-T_0) \ , \label{eq:6.7b}
\end{eqnarray}
and where $A_e(u)$ and $\omega$ are the following functions of $u$,
\begin{eqnarray}
A_e (u) =&& 2 \arctan \,
\left[ \left( {1+e \over 1-e} \right)^{1/2} \tan {u\over 2} \right]
\ , \label{eq:6.7c}\\
\omega =&& \omega_0 + k \, A_e (u) \ ,  \label{eq:6.7d}
\end{eqnarray}
and $u$ is the function of $T$ defined by solving the Kepler equation
\begin{equation}
u - e \sin u = 2 \pi
\left[ \left( {T-T_0 \over P_b} \right)
- {1\over 2} \dot P_b \left( {T-T_0 \over P_b} \right)^2 \right] \ .
\label{eq:6.7e}
\end{equation}
Although the splitting of $F(T)$ into the various contributions
(\ref{eq:6.6b})--(\ref{eq:6.6e}) is a coordinate-dependent concept, one
can loosely say that $\Delta_R$ represents the time of flight across the
(relativistic) orbit (``Roemer time delay"), $\Delta_E$ represents the
combined gravitational and transverse-Doppler redshifts of the pulsar
clock (``Einstein delay"), $\Delta_S$ the gravitational time delay of
the electromagnetic signal propagating in the gravitational potential
generated by the companion (``Shapiro time delay"), while $\Delta_A$ is
associated with aberration effects.

Each theory of gravity makes specific predictions about how the various
parameters (\ref{eq:6.3})--(\ref{eq:6.5}) are related between themselves,
as well as to the (a priori unknown) masses of the pulsar and its
companion.  But the essence of the phenomenological analysis of pulsar
timing data (the so called ``parametrized post-Keplerian" formalism) is
to a priori ignore the existence of such theoretical relations, and to
(least-squares) fit the experimental data to the formula (\ref{eq:6.1}).
The net result of this fit will be to extract in a phenomenological
manner from binary pulsar data the Keplerian parameters (\ref{eq:6.3}),
together with the 8 independent post-Keplerian parameters
(\ref{eq:6.4}).  [For simplicity, we skip the discussion of the fate of
the parameters (\ref{eq:6.5})].

This approach has been generalized to the other pulsar data, those
concerning the shape and polarization of the successive pulses.
Namely, one can write analogs of the formula (\ref{eq:6.1}) for the
pulsar-phase dependence of the observed flux density $S_{\rm obs}
(\nu_{\rm obs}, \phi)$ and linear polarization angle $\psi (\phi)$:
\begin{equation}
S_{\rm obs} (\nu_{\rm obs}, \phi) = G [\phi\,;\, \{p^{\rm K}\}\,;\,
\{\tilde p^{\rm PK}\} ]\ , \label{eq:6.8}
\end{equation}
\begin{equation}
\psi (\phi) = H [\phi\,;\, \{p^{\rm K}\}\,;\,
\{\tilde p^{\rm PK}\} ]\ , \label{eq:6.9}
\end{equation}
where $\phi$ is the rotational phase of the pulsar, and where
\begin{equation}
\{ \tilde p^{\rm PK} \} = \{ \lambda, \dot\lambda, \kappa, \dot\kappa,
\sigma, \dot\sigma, \psi_0, \kappa', \dot\kappa', \sigma', \dot\sigma'
\} \label{eq:6.10}
\end{equation}
is a new set of post-Keplerian parameters, extractable in principle
from pulse structure data.

Summarizing, the parametrized post-Keplerian approach shows that,
besides the easily measured Keplerian parameters (\ref{eq:6.3}), up
to 19 observable post-Keplerian parameters listed in eqs.~(\ref{eq:6.4})
and (\ref{eq:6.10}) can be extracted in a phenomenological manner from
binary pulsar measurements. Any theory of gravity will predict some
specific relations linking these post-Keplerian parameters to the
Keplerian ones, to the masses $m_1$ and $m_2$ of the pulsar and its
companion, and to the Euler angles $\lambda, \eta$ of the pulsar spin
axis. In each theory of gravity we can use 4 of the phenomenological
observables to deduce the values of $m_1$, $m_2$, $\lambda$ and $\eta$,
so that the redundant 15  post-Keplerian 
observables give us 15 tests of the
relativistic law of gravitation.

\subsection{Theory-space approach to binary pulsar tests: introduction
of strong-field parameters $\beta_2$, $\beta'$, $\beta''$,\dots}
\label{sec:6.3}

What is the theoretical significance of the 15 possible tests obtained
by combining measurements of phenomenological parameters~?  What are
these tests teaching us about gravity, and especially about strong-field
and/or radiative aspects of gravity~?  To answer these questions, it is
necessary to generalize to the strong-field regime the alternative-theory
approach discussed above in the quasi-stationary-weak-field context of
the solar system tests.  Fortunately, the same class of
tensor-multi-scalar theories can be used to define a strong-field and
radiative contrast to general relativity.

This was first pointed out in the context of the original
Jordan-Fierz-Brans-Dicke theory. There it was shown that strong
relativistic internal gravitational fields could modify the orbital
dynamics already at the ``Keplerian" level, and generate an a priori
strong emission of dipolar scalar waves \cite{E75} (with the observable
consequence of inducing a corresponding orbital period change in a
binary pulsar).  However, the solar-system tests constrain already so
much the only free parameter of the Jordan-Fierz-Brans-Dicke theory that
this theory no longer provides a sufficient contrast to Einstein's
theory even for what concerns strong-field induced dipole radiation
effects in a system like PSR~1913+16.  [Recently, it was suggested that
the 11-minute binary X-ray source 4U1820-30 could provide a better
testing ground \cite{WZ89}.  However, this system is likely to be
perturbed by the gravitational field of the globular cluster in which it
resides at a level which prevents one from using the observed orbital
period change as a test of gravitation theories; see the update in
Ref.~\cite{W81};  see also Ref.~\cite{G92}].

Recently, the predictions of the most general class of
tensor-multi-scalar theories (containing several arbitrary functions)
have been worked out in detail, with special emphasis on the effects
of strong relativistic internal gravitational fields on the orbital
motion and gravitational radiation reaction in systems of neutron stars
\cite{DEF1}.  One of the main results of this study has been the finding
that there existed (under some assumptions) a strong-field analog of the
weak-field theory parameters $\gamma$ and $\beta$ introduced by Eddington.
More precisely, one finds that, when expanding in powers of the
fractional self-gravity $s=-E^{\rm grav}/mc^2$ all the strong-field and
radiative effects in binary pulsars, the coefficients of these
expansions depend on an infinite sequence of theory parameters
\begin{equation}
\gamma_1, \beta_1, \beta_2, \beta', \beta'', \beta_3, (\beta\beta'),
\ldots \label{eq:6.11}
\end{equation}
All the parameters (\ref{eq:6.11}) are explicitly calculable in terms
of the arbitrary elements entering the action (\ref{eq:5.10}). For
instance (the subscript 0 indicating that an expression is evaluated at
the present cosmological values $\varphi^a_0$)
\begin{eqnarray}
\gamma_1 &=& (\alpha^a \gamma_{ab} \alpha^b)_0 \ , \label{eq:6.12a}\\
\beta_1 &=& (\alpha^a \beta_{ab} \alpha^b)_0 \ , \label{eq:6.12b}\\
\beta_2 &=& (\alpha^a\beta^b_a\beta_{bc}\alpha^c)_0\ ,\label{eq:6.12c}\\
\beta'&=&(\alpha^a\alpha^b\alpha^c D_a\beta_{bc})_0\ , \label{eq:6.12d}\\
\beta'' &=& (\alpha^a \alpha^b \alpha^c \alpha^d D_{ab} \beta_{cd})_0 \ .
\label{eq:6.12e}
\end{eqnarray}
The first two parameters (\ref{eq:6.12a}),(\ref{eq:6.12b}) are equivalent
to the weak-field parameters $\gamma$ and $\beta$ [see
eqs.~(\ref{eq:5.34}) and (\ref{eq:5.37})].  The further parameters
$\beta_2$, $\beta'$, $\beta''$, $\beta_3$,\dots parametrize deeper
layers of structure of the relativistic gravitational interaction which
have been left unprobed by solar system tests.  [As was said above, the
second layer, $\beta_2$, $\beta'$ could be probed by solar system tests
reaching the second-post-Newtonian level of weak-field gravity;  see
\cite{DEF4}].  In pictorial language, each parameter in the list
(\ref{eq:6.11}) represents an (a priori) independent direction away from
general relativity in the space of tensor-multi-scalar theories of
gravity.  The ``post-PPN" parameters $\beta_2$, $\beta'$, $\beta''$,
$\beta_3$,\dots provide a chart for the yet essentially unexplored
domain of strong-gravitational field effects (both in the motion and the
gravitational radiation of systems of strongly self-gravitating bodies).

To give a feeling for the physical significance of the strong-field
parameters (\ref{eq:6.11}) let us mention that the Lagrangian describing
the motion of $N$ strongly self-gravitating bodies [at the approximation
where one treats exactly the strong self-gravitational effects, but
works perturbatively in the inter-body gravitational potential $GM/c^2
({\rm distance}) \sim (v^{\rm orbital}/c)^2$] can be written in the same
form (\ref{eq:5.30}), (\ref{eq:5.32a}) as above if one replaces $G$ by an
effective, body-dependent gravitational constant $G_{AB}$, and similarly
$\gamma$ by $\gamma_{AB}$ and $\beta$ by $\beta^A_{BC}$ [$A,B,
C=1,\ldots,N$ being body labels;  after having done these replacements
one should discard the term $(4\beta -\gamma -3)$ $(E^{\rm grav}_A /m_A
c^2 + E^{\rm grav}_B /m_B c^2)$ which is taken into account in the
replacement $G\to G_{AB}$].  The quantities $G_{AB}$, $\gamma_{AB}$,
$\beta_{BC}^A$ depend on the strength of the self-gravity of the bodies
$A,B$ and $C$.  For instance, if one expands $G_{AB}$ in powers of the
self-gravities of $A$ and $B$ one finds
\begin{eqnarray}
{G_{AB}\over G} = && 1 -{1\over 2}\eta (c_A +c_B) + \left[ \eta +
{\gamma -1\over 2} +\left( {\gamma + 1\over 2}\right)^3 \beta_2
\right] c_A c_B \nonumber \\
&& +(\gamma +1)(\beta -1) (a_A +a_B)
+ \left[ \left( {\gamma +1\over 2}\right)^3
\left(\beta_2 + {1\over 2} \beta'\right)
- 8 (\beta -1)^2\right] (b_A +b_B)
\nonumber \\
&& + O(s^3) \ . \label{eq:6.13}
\end{eqnarray}
In eq. (\ref{eq:6.13}), $\gamma$ and $\beta$ denote the usual weak-field
parameters, $\eta$ denotes the combination $4\beta -\gamma -3$, while
$c_A, a_A, b_A$ denote some ``compactness" factors of body $A$ which are
of order $c_A = -2\partial \ln m_A /\partial \ln G \simeq -2 E^{\rm Grav}
_A /m_A c^2 = O(s_A)$, $a_A =O(s^2_A)$, $b_A =O(s^2_A)$.
Eq.~(\ref{eq:6.13}) shows how the new, strong-field parameters $\beta_2$
and $\beta'$ appear at order $s^2$. Similar formulas exist for the
self-gravity expansions of $\gamma_{AB}$ and $\beta^A_{BC}$. The
strong-self-gravity effects in the gravitational radiation emission
have also been worked out in detail and shown to depend upon $G_{AB}$,
$\gamma_{AB}$, $\beta^A_{BC}$ and a new quantity named $(\alpha_A \beta_B
\beta_C \alpha_D)$. Numerically, the compactness $c_A$ are of order
0.3 for 1.4~$M_\odot$ neutron stars, to be compared to a maximal
compactness $c_A=1$ for black holes. This is a reason for expecting
that self-gravity expansions such as (\ref{eq:6.13}) are useful, even
in the strong-self-gravity context of neutron stars. [See however below].

\subsection{Experimental constraints on strong-field relativistic
gravity.} \label{sec:6.4}

Among the $\sim$ 700 known pulsars, only the class of ($\sim$~35)
``recycled" (millisecond) pulsars furnishes us with potential
relativistic laboratories.  Among the latter, apart from exceptional
cases (notably PSR~1855+09 discussed below), only the subclass of
short-orbital-period, high-eccentricity binary pulsars with neutron
star companions provides interesting gravitational laboratories.  At
present the latter subclass contains only two useful systems:
PSR~1913+16 discovered by Taylor and Hulse in 1974 and PSR~1534+12,
discovered by Wolszczan in 1990 \cite{W91}.

Up to now the phenomenological analysis of the PSR~1913+16 data
has led to the measurement of only 3 post-Keplerian parameters
\cite{TW89}:  $k\equiv \dot\omega P_b /2\pi$ linked to the periastron
advance [eq.~(\ref{eq:6.7d})] $\gamma$ linked to the gravitational redshift of
the pulsar clock [eq.~(\ref{eq:6.6c})], and $\dot P_b$
[eq.~(\ref{eq:6.7e})], the secular change of the orbital period.  In
general relativity, these 3 quantities are predicted to be the following
functions of the masses $m_1$ and $m_2$ of the pulsar and its companion,
\begin{eqnarray}
\dot\omega^{\rm GR} (m_1,m_2)  &=&
{3n\over 1-e^2}\,\left( {GMn\over c^3}\right)^{2/3}\ ,\label{eq:6.14} \\
\gamma^{\rm GR} (m_1,m_2)  &=&    {e\over n}\, X_2\, (1 + X_2)\,
\left( {GMn\over c^3} \right)^{2/3}\ , \label{eq:6.15} \\
\dot P_b^{\rm GR} (m_1,m_2) &=&
-\, {192\pi \over 5c^5}\, X_1 X_2\, (GMn)^{5/3}\,
{ P_4(e)\over (1-e^2)^{7/2}}\ . \label{eq:6.16}
\end{eqnarray}
where we have denoted
\[  M\,\equiv\, m_1 + m_2,\ X_1 \equiv \, m_1/M , \
X_2 \equiv\,m_2 / M \equiv 1- X_1 , \]
\[  n \,\equiv\, 2\pi /P_b, \ P_4(e)\,\equiv\,1+\, {73\over 24}\, e^2 +\,
{37\over 96}\, e^4 \ . \]
The theoretical prediction (\ref{eq:6.16}) for the orbital period
change comes from studying the secular effets of gravitational radiation
reaction in a binary system of two strongly self-gravitating
bodies \cite{D83}.

In graphical terms, the simultaneous measurement of the three
post--Keplerian parameters $\dot\omega^{\rm obs}$, $\gamma^{\rm obs}$
and $\dot P_b^{\rm obs}$ defines, when interpreted within the framework
of general relativity, three curves in the $m_1$, $m_2$ plane, defined
by the equations
\begin{eqnarray}
\dot\omega^{\rm GR} (m_1, m_2)&=& \dot\omega^{\rm obs}\ ,
\label{eq:6.17a} \\
\gamma^{\rm GR} (m_1, m_2)&=& \gamma^{\rm obs} \ ,  \label{eq:6.17b} \\
\dot P_b^{\rm GR} (m_1, m_2)&=& \dot P_b^{\rm obs}\ . \label{eq:6.17c}
\end{eqnarray}
[When taking into account the finite accuracy of the measurements these
curves broaden to three strips in the mass plane].  Equations
(\ref{eq:6.14})--(\ref{eq:6.17a}) thereby yield {\it one} test of {\it general
relativity}, according to whether the three curves meet at one point, as
they should.  As is discussed in detail in \cite{DT91,TWDW,T92}, general
relativity passes this test with complete success [at the accuracy level
$3.5\times 10^{-3}$, given by the  
(experimental, plus Galaxy-induced ) width of the $\dot P_b$ strip].

\medskip

This beautiful success raises at the same time some questions. As
$\dot P_b^{\rm GR}$ is physically due to the radiative structure of the
general relativistic gravitational interaction, one is certainly entitled
to view the $\dot\omega -\gamma -\dot P_b$ test as a convincing
experimental evidence for the existence of gravitational radiation.
However, the rigorous derivations of $\dot P_b^{\rm GR}$ show that the
full strong-field structure of general relativity plays also an
essential role in determining the simple (weak-field-like) formula
(\ref{eq:6.16}).  The same remark applies to the two other formulas
(\ref{eq:6.14}) and (\ref{eq:6.15}).  This is precisely because of this
entangling of various structures of relativistic gravity that it is
useful to analyze the $\dot\omega -\gamma -\dot P_b$ PSR~1913+16 test
within the more general theory-space approach.  This analysis shows
that, in spite of its impressive accuracy, this test can be passed by
theories that deviate significantly from general relativity.

Fortunately, the recently discovered binary pulsar PSR~1534+12 gives us
an independent handle on strong-field gravity. The phenomenological
(parametrized post-Keplerian) analysis of the PSR~1534+12 data allowed
one to extract 4 independent post-Keplerian parameters: $\dot\omega$,
$\gamma$, $r$ and $s$. [The latter two entering the gravitational time
delay (\ref{eq:6.6d})]. Within each theory of gravity $\dot\omega$,
$\gamma$, $r$ and $s$ are predicted to be some specific functions of
the two masses $m_1$ and $m_2$. Therefore these 4 phenomenological
measurements define 4 curves in the $(m_1, m_2)$ plane of the masses of
PSR~1534+12 and its companion (beware that this is a different mass plane
than the one associated to PSR~1913+16).  This means that we thereby get
$4-2=2$ tests of any theory of gravity, according to whether the four
curves meet at one point.  One finds that general relativity passes
these two new tests with complete success.  It is important to note that
these tests concern the quasi-stationary, strong-field regime without
mixing of radiative effects.  At present, the accuracy of these
strong-field tests is not very high, but numerical simulations show that
they should steadily improve as more data become available.  The system
PSR~1534+12 may offer also the possibility of seeing (for the first
time) the relativistic spin-precession induced by the gravitational
spin-orbit coupling [through a careful monitoring of the secular changes
of the pulse shape].  Indeed, in this system (contrary to PSR~1913+16)
the spin axis is significantly misaligned with the orbital angular
momentum (by at least 8$^\circ$) \cite{DT92}. Recently, it has also been
possible to measure (at the 20\% precision level) ${\dot P}_b$ in PSR1534+12,
with a result in agreement with general relativity \cite{Nobel}.

Shifting from the phenomenological to the theory-space approach, one
can ask to what extent all the existing pulsar data constrain the
possible relativistic theories of gravity, beyond the solar-system data.
This question has been recently addressed, using as space of theories a
specific two-parameter class of tensor-bi-scalar theories, called
$T(\beta', \beta'')$.  This class was introduced in Ref.~\cite{DEF1} to
describe the two yet unexplored directions in theory space associated
with the strong-field parameters $\beta'$ and $\beta''$, independently
of the already explored directions (i.e.  independently from the
weak-field directions $\gamma$ and $\beta$, and from the strong-field
dipole radiation effects explored in Refs.~\cite{E75,WZ89}).

One made use of (i) 10 years of high-quality timing observations of
PSR~1913+16, (ii), one year of similar data for PSR~1534+12, and (ii)
a previously published constraint on a possible strong-field violation
of the strong equivalence principle \cite{DS91}, based on an
interpretation of the data of the ``non relativistic" binary pulsar
PSR~1855+09 \cite{RT91}.  Each set of data selects within the
two-dimensional plane of theories $(\beta', \beta'')$ some allowed
region [Specifically, the region where the $\chi^2$ statistics,
eq.~(\ref{eq:2.2}), is smaller than the level corresponding to a
(formal) 90~\% confidence level].  In this way, one gets the following
three regions in the $\beta'$, $\beta''$ plane (see Fig.3 in 
ref. \cite{TWDW}):  (i) a thin strip,
roughly located around the parabola $\beta''= (\beta')^2$, corresponding
to the single (0.5~\% accurate) $\dot\omega -\gamma -\dot P_b$ 1913+16
test, (ii) a wide potato-shaped region corresponding to the two new
(low-precision) $\dot\omega -\gamma -r-s$ 1534+12 tests, and (iii) the
vertical strip $-1.6 < \beta' < +1.5$ corresponding to the $e-P_b$
1855+09 test.  When combining these three independent allowed regions in
theory space one gets two interesting results:

(1) the three allowed regions do admit a non empty common intersection,
and general relativity [i.e. the point $(\beta', \beta'')=(0,0)]$ lies
well inside this intersection region,

(2) at the 90~\% confidence level the theory parameters $\beta'$ and
$\beta''$ are constrained to lie in a thin parabolic segment whose
projections on the $\beta' \beta''$ axes are roughly $-1.1 <\beta' <1.6$,
$-1 <\beta'' <6.$

>From a quantitative point of view, these limits are less impressive than
the ones obtained on the weak-field parameters $\gamma$ and $\beta$ by
using solar-system data. However, they represent our first limits on
possible strong-field effects in the motion and radiation of systems of
neutron stars.  One should note also that, from a theoretical point of
view, the $T(\beta', \beta'')$ class of tensor-bi-scalar theories
considered in the previous analysis has the unpleasing feature of
containing ghost (i.e. negative-energy) excitations. [A feature actually
shared by all previously considered strong-field alternatives to general
relativity, except the uninteresting, because already too constrained,
Jordan-Fierz-Brans-Dicke theory].  It was thought in Ref.~\cite{DEF1}
that the presence of ghosts (i.e.  the indefiniteness of the
$\sigma$-model metric $\gamma_{ab}$) was necessary to construct a class
of theories obeying the tight weak-field limits (\ref{eq:5.40}),
(\ref{eq:5.43}), and still exhibiting significant strong-field
departures from general relativity.  [This is linked to the discussion
above of one's theoretical pessimism concerning possible 2PN deviations,
given the existing tight 1PN limits].  Actually, it has been recently
discovered \cite{DEF3} that a physically fully satisfactory class of
tensor-scalar theories (containing only positive-energy excitations, and
satisfying the weak-field tests) exhibited non-perturbative strong-field
effects (showing up when considering the exact, infinite series of
self-gravity effects) which allowed strong-field departures from general
relativity.  The extent to which actual binary pulsar data, (and
cosmological considerations), constrain these appealing strong-field
alternatives to general relativity is presently being investigated.

\section{Cosmology}\label{sec:7}

\subsection{Introduction}\label{sec:7.1}

All the tests considered above have examined the gravitational interaction
on scales between 1~mm and a few astronomical units (1AU$\simeq
1.5\times 10^8$~km). See also \cite{W87} for a discussion of astrophysical
tests that we do not consider here. In principle, the Universe is providing us
with plenty of data concerning the behaviour of gravity on large scales. However,
most of these data cannot be used as clean tests of the law of gravity because
of our lack of a priori knowledge of the matter distribution, and/or the low
accuracy of the data themselves (especially in certain cosmological data).  For
instance, there are well established cases (rotation curves in the outer regions
of many spiral galaxies, velocity dispersions in some clusters of galaxies)
where there is a significant discrepancy between the mass that we can infer from
the observed light and the mass needed to hold the system in gravitational
equilibrium if Newton's law is assumed to be valid.  This discrepancy
needs not indicate that Newton's faw is at fault (see, however, \cite{MOND},
\cite{BBS91}) because there may well be a lot of unseen (``dark") matter in these
systems.  For reviews on this ``dark matter" issue see Ref.~\cite{DM}. 
Fortunately, there are a few cases where one can factor out one's ignorance of
the real matter distribution and get rather direct tests of the validity of
general relativity on large scales.  This happens in particular in some cases of
gravitational lensing of distant quasars or galaxies by intermediate
galaxies or clusters of galaxies (For a review on gravitational lensing
see e.g.  Ref.~\cite{SEF}).  For instance one observes a giant optical
arc near the center of the rich cluster of galaxies A370.  This arc is
an optical mirage, coming from the lensing of the light of a distant
galaxy by the gravitational field of the intermediate cluster.  The
radius of curvature of the arc, $\theta =26'' \pm 2''$, is therefore a
quantitative way of probing the gravitational field of the cluster.  On
the other hand, the dispersion of the velocities of the galaxies making
up the cluster, $\sigma =1300 -1700$~km/s, gives us another quantitative
probe of the mean gravitational field of the cluster.  One can verify
whether the general relativistic prediction linking $\theta$ to
$\sigma^2$ (and to the redshifts of the distant and intermediate
galaxies) is satisfied, without having to know in advance the real
matter distribution in the cluster.  One finds that the test is
satisfied within a precision of order 30~\%.  Therefore, within this
precision one has verified the general relativistic action on light and
matter of an external gravitational field on a length scale
$\sim$~100~kiloparsec.  See Ref.~\cite{Dar} for a discussion of the use
of A370 and other simple cases of gravitational lensing as tests of
general relativity at large distances.  The typical accuracy of these
tests is $\sim$~30\%.

Let us now turn our attention to the constraints on gravitation coming from
cosmological data. A first type of constraints comes from considering that
our universe certainly went through a very hot and dense phase (``hot big
bang").  The observation of the cosmological microwave background [isotropy and
black body spectrum] establishes that the temperature of the universe once
exceeded $4\times 10^3$~K (temperature of ionization of hydrogen).  Even if one
doubts the details of primordial nucleosynthesis, it seems pretty certain that
most of the helium in the universe was formed during the early big bang, when the
temperature exceeded $10^9$~K.  With less confidence, one can use the comparison
between the big bang computations of the abundances of light elements
(essentially He$^4$, with traces of H$^2$, H$^3$ and Li$^7$) and the present
observations to set constraints on the gravitation theory ruling the evolution
of the early universe \cite{GS76,B78,Y79}. [One must however keep in mind the
large uncertainties brought by the need to extrapolate the observed abundances
back in time, as well as the possibility of inhomogeneities in the early
universe \cite{VG84,V93}]. Basically, this comparison is constraining the rate
of expansion of the universe during nucleosynthesis.  If we assume that the
number of degrees of freedom of the thermalized matter of the hot big bang around
$10^{10} - 10^9$K is that given by the Standard Model (with three light
neutrinos), the nucleosynthesis constraint is essentially giving us a
constraint on the value of the gravitational coupling constant during
nucleosynthesis. One obtain limits of order $| \Delta G/G | \lessim 10\%$ (see
e.g. \cite{RM82} and references therein).

In principle, one can also constrain gravitation theories by combining
measurements of the present expansion rate $H_0$ and of the age of
the universe $T_0$. For instance, it has been argued that the product
$H_0T_0$ was greater than 0.4 at the ``95~\% C.L." \cite{F87}.  However,
the large systematic uncertainties in the determination of $H_0$ and
$T_0$ make it difficult to assess the significance of such a limit.

At present, general relativity is nicely consistent with all the above
cosmological tests. [Note that, from time to time, the $H_0T_0$ test
has been claimed to present a real problem for general relativity].
As an illustration of the use of the various cosmological tests to
constrain the space of possible theories one may consult
Refs~\cite{W81,RM82,DGG}.

An important issue concerns the possible existence of scalar
fields having only gravitational-strength couplings to matter.
For instance, the ``gravitational sector'' of string theory contains, besides
the standard Einsteinian tensor field $g_{\mu \nu}$, some gauge-neutral scalar
fields with Planck-suppressed couplings. We refer to these fields as moduli
(they include the model-independent dilaton $S \sim e^{-2\Phi}$). The moduli
fields are massless to all orders of perturbation theory and play a central
physical role in string theory in that their vacuum expectation values (VEV)
determine the coupling constants of the theory: notably the string
coupling constant $g_{\rm string} = e^{\Phi}$ (associated
to the string loop expansion), the (unified) gauge
coupling constant $g_{\rm gauge}^{-2} = S + f(T) + \cdots$
\cite{DKL} and the gravitational constant (in string units)
$G \sim \alpha' \, [e^{2\Phi} + \cdots ]$. These crucial
properties of the moduli suggest that they may play an
important role in cosmology and in low-energy gravity.

\subsection{Cosmology of moduli fields}\label{sec:7.2}

Let us recall the basic equations of homogeneous Friedmann
cosmology (in an ``Einstein'' conformal frame, i.e. with
standard kinetic terms for $g_{\mu \nu}$)
\begin{equation}
ds^2 = g_{\mu \nu} \, dx^{\mu} \, dx^{\nu} = -dt^2 + a^2
(t) \, \left[ {dr^2 \over 1-kr^2} + r^2  d\theta^2 +
r^2  \sin^2 \theta d\varphi^2 \right] \ , \label{eq:123}
\end{equation}
\begin{equation}
\left( {\dot a \over a}\right)^2 + {k\over a^2} = {8\pi G
\over 3} \, \rho \ , \label{eq:124}
\end{equation}
\begin{equation}
{\ddot a \over a} = -{4\pi G \over 3} \, (\rho +3p) \ . \label{eq:125}
\end{equation}
The source terms $\rho$ and $p$ satisfy the energy balance
law
\begin{equation}
{d\over dt} \, (\rho \, a^3) = -p \, {da^3 \over dt} \ , \label{eq:126}
\end{equation}
and are given by a sum of zero-pressure (non-relativistic
matter), radiative, field-kinetic-energy and potential
contributions. In the case of one modulus with standard
kinetic term:
\begin{equation}
\rho = \rho_{\rm nr} + \rho_{\rm rad} + {\textstyle {1\over
2}} \, {\dot \varphi}^2 + V(\varphi) \ , \label{eq:127}
\end{equation}
\begin{equation}
p = 0 + {\textstyle {1\over 3}} \, \rho_{\rm rad} +
{\textstyle {1\over 2}} \, {\dot \varphi}^2 - V(\varphi) \ . \label{eq:128}
\end{equation}
When one type of matter contribution dominates, the
pressure over density ratio $\lambda \equiv p/ \rho$ is
approximately constant $(\lambda_{\rm nr} = 0, \
\lambda_{\rm rad} = {1\over 3} , \ \lambda_{\rm kinetic} =
+1, \ \lambda_{\rm potential} =-1)$ and the energy density
varies with the scale factor as $\rho \propto a^{-3
(1+\lambda)}$. This gives, respectively, $a^{-3}$,
$a^{-4}$, $a^{-6}$ and $a^0$ in the
non-relativistic-, radiation-, kinetic-energy- and
potential-driven expansion. The corresponding expansions
follow a power-law, $a(t) \propto t^{{2\over
3(1+\lambda)}}$, when $\lambda \not= -1$, and an
exponential law when $\lambda =-1$ (potential-driven
inflation). [In terms of the conformal time $\eta = \int
dt / a(t)$, this becomes $a(\eta) \propto \eta^{{2\over
1+3\lambda}}$.] As there is always some thermal radiation
around, note that a kinetic-energy driven expansion is
unstable and necessarily becomes rapidly dominated by
radiation. [This was an argument levelled long ago by
Grishchuk against the proposal of Zeldovich to use the
``hard'' equation of state $p=\rho$ to describe the early
universe.]

Let us also recall some of the puzzling features of our
present large-scale universe that we would like to explain
in a natural manner:
\begin{enumerate}
\item[a.] The extreme smallness of the vacuum energy
(cosmological constant) on any a priori relevant
particle-physics mass scale: 
\begin{equation}
\rho_{\rm vac} \lessim {3\over 8\pi G} \, H_0^2 \sim
10^{-120} \, m_P^4 \sim 10^{-41} \, \Lambda_{\rm QCD}^4 \label{eq:129}
\end{equation}
(where $m_P \equiv G^{-1/2}$ denotes the Planck mass).
\item[b.] The fact that our universe has been expanding for
a time $\grtsim 10^{10} \, {\rm yr} \sim 10^{61} \,
m_P^{-1}$ without either recollapsing or becoming dominated
by a spatially negative curvature. This implies that early
on the space  curvature term $k/a^2$ in Eq. (\ref{eq:124}) was
negligible compared to the ``time curvature'' term $H^2
\equiv (\dot a /a)^2$.
\item[c.] The extreme homogeneity of the universe over
$\sim 10^5$ causally disconnected regions at the time of
last scattering of the cosmic microwave background.
\end{enumerate}

The inflationary scenarios give a physical explanation of
the facts b. and c., but always at the price of some
fine-tuning of parameters.

Binetruy and Gaillard \cite{BG86} were among the first to
ask whether moduli could be useful in cosmology. They tried to see whether any
of the moduli fields could provide a natural candidate for being an
``inflaton'', i.e. for driving a sufficiently long stage of exponential inflation
through the dominance of its potential energy. They did
not find any natural candidate for the inflaton among the
moduli. Later work, notably one by Campbell, Linde and
Olive \cite{CLO}, stressed the specific obstacles to a
successful inflationary scenario brought by the existence
of the dilaton. In particular, instead of driving an
exponential inflationary expansion, a (string-frame)
constant energy density drives the dilaton towards large
negative values (corresponding to weak couplings), while
the universe expands only as a small power of time.
Another problem linked to the dilaton is the shallowness
of the nonperturbative potentials it might acquire
\cite{BS93}. This makes it difficult to see how (without
fine-tuning the initial conditions) its potential can fix
the VEV of the dilaton at a reasonable value.

On the other hand, some interesting, qualitatively new,
features of string cosmology have been explored
\cite{ABEN}, \cite{TV92}, \cite{V91}, \cite{GV93}. In
particular, Veneziano and Gasperini \cite{V91},
\cite{GV93}, motivated by the ``scale-factor duality'' of
the tree-level string effective action
\begin{equation}
S = \int d^4 x \, \sqrt{\hat g} \, e^{-2\Phi} \, \left[
\hat R (\hat g ) + 4 (\hat{\nabla} \, \Phi)^2 \right] \label{eq:130}
\end{equation}
(i.e. the symmetry $\hat a (t) \rightarrow \hat{a}^{-1} \,
(t)$, $\Phi (t) \rightarrow \Phi (t) - 3 \, {\rm ln} \,
\hat a (t)$), introduced a ``pre-big-bang scenario'' in
which our present stage of decelerated expansion was
preceded by a ``super-inflationary'' stage of accelerated
expansion driven by the kinetic energy of the dilaton. In
the string conformal frame (i.e. using the $\sigma$-model
metric $\hat{g}_{\mu \nu}$), this pre-big-bang solution
reads $(\hat t < 0)$
\begin{equation}
\hat a \, \propto (-\hat t )^{-1/\sqrt 3} \ , \ \Phi =-
{1 + \sqrt 3 \over 2} \, {\rm ln} (-\hat t ) \ , \label{eq:131}
\end{equation}
while, in the Einstein frame (metric $g_{\mu \nu} =
e^{-2\Phi} \, \hat{g}_{\mu \nu}$) it corresponds to an
accelerated {\it contraction}
\begin{equation}
a \, \propto (-t)^{1/3} \ , \ \Phi = -{1\over \sqrt 3} \,
{\rm ln} (-t) \ . \label{eq:132}
\end{equation}
[Note that the fact that it contracts in Einstein units
ensures the stability of this kinetic-energy driven
dynamics against the unavoidable presence of thermal
radiation.] This scenario can provide a large amount of
inflation and several of its possible observational
consequences have been discussed: gravitational waves,
relic dilatons, generation of primordial galactic magnetic
fields \cite{GV94}. The main shortcoming of this scenario
is that it postulates, without being able to describe, the
existence of a strong-curvature transition between the
pre-big-bang stage (\ref{eq:131}) and the standard Friedmann-Gamow
hot big bang.

On a less ambitious vein, one can ask whether the presence
of moduli, i.e. the existence of scalar fields with
Planck-scale natural range of variation, can help in
solving the endemic fine-tuning problems of
potential-driven inflationary scenarios. Let us consider
the action describing the dynamics of gravity (described
in the Einstein frame) and an arbitrary number of moduli
fields having a potential $V(\varphi^a)$ $[a=1,\ldots ,n]$
\begin{equation}
S = \int d^4 x \, \sqrt g \, \left\{ {\tilde{m}_P^2 \over
4} \, R (g_{\mu \nu}) - {\tilde{m}_P^2 \over 2} \,
\gamma_{ab} (\varphi^c) \, g^{\mu \nu} \, \partial_{\mu}
\, \varphi^a \, \partial_{\nu} \, \varphi^b - V(\varphi^a)
\right\} \ . \label{eq:133}
\end{equation}
Here $\tilde{m}_P = m_P / \sqrt{4\pi} = (4\pi G)^{-1/2}$
is a reduced Planck mass, and the fields $\varphi^a$ are
dimensionless. Following \cite{DN92}, it is useful to
combine the Friedmann equations for the scale factor
$a(t)$ with the equations of motion of the moduli
$\varphi^a (t)$ to write an {\it autonomous} equation
describing the evolution of the $\varphi$'s in terms of
the parameter
\begin{equation}
p = \int H \, dt = \int {\dot a \over a} \, dt = {\rm ln} \,
a + \hbox{const} \label{eq:134}
\end{equation}
measuring the number of $e$-folds of the expansion. This
yields (when $k=0$) the simple-looking equation
\begin{equation}
{2\over 3- {\varphi \!\!\!\! \varphi \!\!\!\!
\varphi}^{'2}} \, {\varphi \!\!\!\! \varphi
\!\!\!\! \varphi}''_{\rm cov} + 2 \, {\varphi \!\!\!\!
\varphi \!\!\!\! \varphi}' = - \nabla_{\varphi \!\!\!\!
\varphi \!\!\!\! \varphi} \ {\rm ln} \vert V ({ \varphi
\!\!\!\! \varphi \!\!\!\! \varphi})\vert \ , \label{eq:135}
\end{equation}
where ${\varphi \!\!\!\! \varphi \!\!\!\! \varphi}' \equiv
d \, {\varphi \!\!\!\! \varphi \!\!\!\! \varphi} / dp$ and
where ${\varphi \!\!\!\! \varphi \!\!\!\! \varphi}''_{\rm
cov}$ denotes the covariant derivative of 
${\varphi \!\!\!\! \varphi \!\!\!\! \varphi}'$ with
respect to the $\sigma$-model metric $\gamma_{ab} (
{\varphi \!\!\!\! \varphi \!\!\!\! \varphi}) \, d\varphi^a
\, d\varphi^b$.

The generic solution of Eq. (\ref{eq:135}) can be easily grasped
from a simple mechanical analogy: a particle with position
${\varphi \!\!\!\! \varphi \!\!\!\! \varphi}$ and
velocity-dependent mass $m({\varphi \!\!\!\! \varphi 
\!\!\!\! \varphi}') = 2/(3- {\varphi \!\!\!\! \varphi 
\!\!\!\! \varphi}^{'2})$ moves, in $p$-time, in a curved
manifold submitted to the external potential ${\rm ln}
\vert V({\varphi \!\!\!\! \varphi \!\!\!\! \varphi})\vert$
and a constant friction $-2 {\varphi \!\!\!\! \varphi 
\!\!\!\! \varphi}'$. If the {\it curvature} of the
effective potential ${\rm ln} \, V({\varphi \!\!\!\! 
\varphi \!\!\!\! \varphi})$ is sufficiently small, more
precisely if, in the one-scalar case,
\begin{equation}
{\partial_{\varphi}^2 \, {\rm ln} \, V \over 6 - {1\over
2} \, (\partial_{\varphi} \, {\rm ln} \, V)^2} \ll 1 \ , \label{eq:136}
\end{equation}
the motion of ${\varphi \!\!\!\! \varphi \!\!\!\!
\varphi}$ is rapidly friction-dominated:
\begin{equation}
2 \, {d {\varphi \!\!\!\! \varphi \!\!\!\! \varphi} \over
dp} \simeq - \nabla_{{\varphi \!\!\!\! \varphi \!\!\!\!
\varphi}} \ {\rm ln} \, V({\varphi \!\!\!\! \varphi
 \!\!\!\! \varphi}) \ . \label{eq:137}
\end{equation}

In the one-scalar case, Eq. (\ref{eq:137}) directly gives the number
of $e$-folds as a function of the ``inflaton'' $\varphi$
\begin{equation}
N = \int dp \simeq \int {2 \, d\varphi \over
\partial_{\varphi} \, {\rm ln} \, V} \ . \label{eq:138}
\end{equation}
Note that the ``scale of inflation'' $m_I$, such that
\begin{equation}
V(\varphi) = m_I^4 \, v(\varphi) \ , \label{eq:139}
\end{equation}
drops out completely from equations (\ref{eq:135})-(\ref{eq:138}). Only
the dimensionless logarithmic shape ${\rm ln} \,
v(\varphi)$ matters. This shape needs to be sufficiently
flat for inflation to continue during $N> 65$. Here, one
finds the first need of a fine tuning: $\partial_{\varphi}
\, {\rm ln} \, v(\varphi) \ll 1$ to ensure
\begin{equation}
N = \int {2 \, d\varphi \over \partial_{\varphi} \, {\rm ln}
\, v(\varphi)} > 65 \ . \label{eq:140}
\end{equation}
Banks et al. \cite{BBSMS} emphasized, however, that it is
relatively favourable to have a modulus field as inflaton,
as a canonical scalar field $\phi$ varying on a typical
range $f< \tilde{m}_P$ (i.e. $V(\phi) = m_I^4 \,
v(\phi/f)$) would imply a number of $e$-folds smaller by a
factor $(f/\tilde{m}_p)^2 : N = (f/\tilde{m}_P)^2 \int 2
d\varphi / \partial_{\varphi} \, {\rm ln} \, v(\varphi)$
[where $\varphi = \phi /f$].

Even when the inflaton is a modulus, it remains to find a
natural explanation for having a large number of
$e$-folds, Eq. (\ref{eq:140}), i.e. for starting the evolution on a
very flat region of ${\rm ln} \, v(\varphi)$. Several
mechanisms have been proposed to this end. For instance:
\cite{SBB89} envisaged potentials levelling off to a
constant value when the (canonical) inflaton takes large
values; \cite{V} (see also \cite{DV95}) argued that
quantum cosmology suggests that universes, spontaneously
nucleating out of nothing, preferably start at a {\it
maximum} of $V(\varphi)$ [because the instanton action
$\vert S \vert = 3m_P^4 / 8 V(\varphi)$ is minimized there];
and \cite{BBSMS} invoked the use of stringy domain walls or
other topological defects \cite{CQR} as a natural
mechanism for triggering inflation \cite{V94},
\cite{LL94}. However, even if one starts the evolution at
the top of a potential barrier, one needs an uncomfortably
small curvature there: if $v(\varphi) \simeq v_{\rm max}
\, \left[ 1-{1\over 2} \, \beta (\varphi -\varphi_m)^2
\right]$ one needs $\beta \lessim 10^{-2}$ \cite{BBSMS}.

It remains then to satisfy the strong constraint that the
density fluctuations generated by inflation be smaller
than about $10^{-5}$. This constraint reads
\begin{equation}
{\delta \rho \over \rho} \sim {\tilde{m}_P \, H_I^3 \over
\partial V / \partial \varphi} \sim \left( {m_I \over
\tilde{m}_P} \right)^2 \, {v^{3/2} (\varphi) \over
\partial_{\varphi} \, v(\varphi)} \lessim 10^{-5} \ . \label{eq:141}
\end{equation}
The other constraint that horizon-wave-length
gravitational waves be compatible with the observed
isotropy of the cosmic microwave background is $h_{GW}
\sim H_I / \tilde{m}_P \sim (m_I / \tilde{m}_P)^2 \lessim
10^{-5}$, and is generically weaker then Eq. (\ref{eq:141}) because
of the necessary flatness of ${\rm ln} \, v(\varphi)$. The
constraint (\ref{eq:141}) creates a mass-scale problem: First, it
seems to exclude that inflation be directly generated at
the string scale (even if $m_{\rm string} \sim 10^{17}
{\rm GeV} < \tilde{m}_P = 3.4 \times 10^{18} \, {\rm GeV}$
\cite{K88}). Second, the inflationary mass scale it
suggests, $m_I \sim 10^{15} \, {\rm GeV}$, seems totally
disconnected from the preferred SUSY breaking scale $m_{\rm
SUSY} \sim \sqrt{m_{3/2} \, \tilde{m}_P} \sim 10^{11} \,
{\rm GeV}$ with $m_{3/2} \sim 1 \, {\rm TeV}$. [Note,
however, that $m_I$ can be $\ll 10^{15} \, {\rm GeV}$ if
cosmic strings are invoked to generate the needed initial
density fluctuations.]

Even if one does not try to use the moduli as inflatons,
the existence of light gravitationally coupled fields
resurrects the infamous Polonyi problem \cite{CFKRR},
\cite{GLV}, \cite{ENQ}, \cite{ETV}. In essence the problem
is that the presence of light particles with very weak,
Planck-suppressed couplings causes cosmological problems,
{\it either} because they decay late and generate too much
entropy while failing to reheat the Universe sufficiently
to restart nucleosynthesis, {\it or} because they do not
decay and overdominate the Universe through the energy
density stored in the oscillations of their zero-mode in
their potential $V(\phi) \simeq {1\over 2} \, m_{\phi}^2
(\phi - \phi_m)^2$. The three crucial parameters at the
root of the problem are: (i) the initial displacement
$\Delta \phi$ of the VEV of $\phi$ away from the minimum
$\phi_m$ of its low-energy potential; (ii) the mass
$m_{\phi}$ of $\phi$; and (iii) the decay rate $\Gamma$ of
$\phi$. The problem arises because of the generally
expected links:
\begin{equation}
\Delta \phi \sim \tilde{m}_P \ , \ \Gamma \sim m_{\phi}^3
/ \tilde{m}_P^2 \ . \label{eq:142}
\end{equation}

When (\ref{eq:142}) hold one finds that the entire range of masses
$10^{-28} \, {\rm eV} \lessim m_{\phi} \lessim 30 \, {\rm
TeV}$ is excluded. More precisely, if $10^{-28} \, {\rm eV}
\lessim m_{\phi} \lessim 100 \, {\rm MeV}$ (so that $\Gamma
\lessim H_0$) the field $\phi$ has not decayed by now and
the energy stored in ${1\over 2} \, \dot{\phi}^2 + {1\over
2} \, m_{\phi}^2 \, \phi^2$ (which decreases $\propto \,
a^{-3}$) overdominates the Universe, while, if $100 \,
{\rm MeV} \lessim m_{\phi} \lessim 30 \, {\rm TeV}$ the
field has decayed by now, but its decay has reheated the
Universe at a temperature $T_R \lessim 1 \, {\rm MeV}$, too
small to restart nucleosynthesis, and has produced an
enormous amount of entropy diluting away the results of any
previous nucleosynthesis. The problem cannot be evaded by
a long period of ordinary inflation as the latter
regenerates via long-wave quantum fluctuations (which are
important for $\phi$ if $m_{\phi} \ll H_I$), an
unacceptably large VEV for $\phi$ \cite{GLV}. The Polonyi
problem is a serious difficulty for {\it all} moduli
because, as stressed in \cite{CCQR}, \cite{BKE},
\cite{BBS}, \cite{BBSMS}, current SUSY breaking lore
suggests that they (as well as their fermionic partners)
acquire masses of order $m_{\phi} \sim m_{3/2} \sim 1 \,
{\rm TeV}$, which is uncomfortably below the $30 \, {\rm
TeV}$ limit mentioned above. [In essence, this mass estimate
follows from $V(\phi) = m_{\rm SUSY}^4 \, v
(\phi/\tilde{m}_P) = m_{3/2}^2 \, \tilde{m}_P^2 \, v(\phi /
\tilde{m}_P)$.]

Some solutions to the modular Polonyi problem have been
proposed. In particular, \cite{RT94} pointed out that the
potential $V(\phi) = m_{\rm SUSY}^4 \, v(\phi /
\tilde{m}_P)$, which is at the origin of the problematic
value $m_{\phi} \sim m_{3/2}$ of $m_{\phi}$, might solve
the problem by generating a brief period of (secondary)
inflation (with one of the moduli as inflaton) at a ``weak
scale'' expansion rate: $H_I \sim V^{1/2} / \tilde{m}_P
\sim m_{\rm SUSY}^2 / \tilde{m}_P \sim m_{3/2}$. They find
that a few $e$-folds of inflation with $H_I \lessim
m_{\phi}$ is enough to sufficiently decrease $\Delta
\phi$, and thereby the energy stored in $V(\phi)$. Another
type of solution has been proposed in \cite{DV95}. This
reference (see also \cite{DRT95}) shows that the cosmological difficulties of the
moduli are avoided if the mechanism introduced by Damour
and Polyakov \cite{40a} for fixing the VEVs of the moduli
is at work.

In brief, the point of \cite{40a} was to show that,
contrary to what is usually assumed, having one or several
of the moduli stay exactly massless in the low-energy
world can be naturally (i.e without fine-tuning of
parameters) compatible with existing experimental data.
Two conditions must be satisfied for this to happen:

(i) string-loops effects must generate a non-trivial
dependence of the moduli-dependent coupling functions
entering the effective Lagrangian, i.e. the universal
multiplicative factor $e^{-2 \Phi}$ entering the
tree-level action (3.1) must get replaced by various
moduli-dependent functions,
\begin{equation}
S = \int d^4 x \, \sqrt{\hat g} \, \left\{ {B_g
(\varphi^a) \over \alpha'} \, \hat R - {B_{ab} (\varphi^c)
\over \alpha'} \, \hat{g}^{\mu \nu} \, \partial_{\mu} \,
\varphi^a \, \partial_{\nu} \, \varphi^b - {k\over 4} \,
B_F (\varphi^a) \, \hat{F}_{\mu \nu}^2 + \cdots \right\} \ , \label{eq:143}
\end{equation}
where $B_i (\varphi^a) = e^{-2\Phi} + f_i (\Phi ,T)$
admits extrema at finite values of the $\varphi$'s;

(ii) there exist some preferred values of the $\varphi$'s,
say $\varphi_m^a$, where all the $B_i$'s relevant to
determining the low-energy mass scales (notably the gauge
coupling function $B_F (\varphi^a)$ which determines
$\Lambda_{\rm QCD} (\varphi)$) reach (at least
approximately) an extremum. A simple mechanism for
ensuring this property might be the existence of a
discrete symmetry ($S$-duality, $T$-duality) in moduli
space. [See \cite{40a} for further suggestions.]

Under these assumptions, \cite{40a} finds that a ``least
coupling principle'' holds in that the cosmological
expansion naturally drives the VEVs of the $\varphi$'s
toward $\varphi_m^a$, where the moduli (classically)
decouple from matter. Estimates of the small, but non
zero, present values of $\varphi -\varphi_m$ show that
they are compatible with existing experimental data,
including the extremely stringent tests of the equivalence
principle ($\sim 10^{-12}$ level). \cite{DV95} has studied
the consequences of this mechanism when considering an
early stage of inflation. They found it fully compatible
with observational facts. In particular, quantum
fluctuations are inefficient in regenerating a
quasi-classical long-wavelength VEV for the $\varphi$'s.
They also pointed out that, if one considers {\it massive}
moduli, the Polonyi problem is naturally avoided in such a
scenario because the moduli acquire during inflation an
effective mass $m_{\varphi} \sim H_I$ so that they are very
efficiently driven to the preferred values $\varphi_m$ at
which they store no potential energy. Moreover, in this
model moduli within a very wide range of masses (which,
contrary to usual models, include the SUSY-breaking
favored $\sim 1 \, {\rm TeV}$ value) qualify to define a
novel type of essentially {\it stable ultra-weakly
interacting dark matter}. Indeed, one finds that quantum
fluctuations generate the following contribution to the
cosmological closure density in the form of massive moduli:
\begin{equation}
\Omega_{\varphi} \sim \left( 10^5 \, {H_I \over m_P}
\right)^{{3\over 2}} \, {m_{\varphi} \over 10 \, {\rm GeV}}
\ . \label{eq:144}
\end{equation}
If the inflationary mass scale $m_I \sim (H_I \,
\tilde{m}_P)^{{1\over 2}} \sim 4 \times 10^{15} \, {\rm
GeV}$, a modulus of mass $m_{\varphi} \sim 1 \, {\rm TeV}$
can close the Universe: $\Omega_{\varphi} \sim 1$.

\section{Conclusions}\label{sec:8}

To complete this review one should mention the fact that gravitational
wave observations in Earth-bound interferometric detectors
(LIGO/VIRGO/\dots) should soon give us access to many new ways of
probing the regime of strong and rapidly varying gravitational fields.
First, simultaneous observations from an array of interferometric
detectors can in principle verify whether the gravitational waves
received on Earth are of the pure massless, helicity-2 type predicted
by general relativity, or contain other excitations. Second, the
amplitude and shape the detected signals will tell us a lot about the
theory of gravity. For instance, if the correct theory of gravity is a
tensor-scalar one presenting the kind of nonperturbative strong-field
behaviour mentioned above, one expects that the spherically symmetric
collapse of the core of a star down to a neutron star state will emit
strong monopolar scalar waves, with energy flux of order
\begin{equation}
{dE^{\rm spin 0 waves} \over dt} \sim {G\over c^5} \,
\left( {d\over dt}\, E^{\rm grav} \right)^2 \ , \label{eq:7.2}
\end{equation}
where $E^{\rm grav}$ is the gravitational binding energy of the
collapsing core.  [See eq. \ref{eq:7.2} of Ref.~\cite{DEF1}].  The flux
(\ref{eq:7.2}) is expected to be much bigger than the corresponding pure
spin 2 result (which depends crucially on the deviations from spherical
symmetry).  However, the detection on Earth of scalar waves will be
hampered by a small factor $\alpha =\sqrt{(1-\gamma)/2} < 0.032$
\cite{DEF1}.  Still, an optimist could hope to detect spin~0
gravitational waves from stellar collapses (and thereby to falsify
Einstein's theory) before being able to detect the spin~2 waves emitted
by the inspiralling motion of binary neutron stars~!

The first general conclusion one can draw from the above review of the
experimental situation is that Einstein's theory of gravity has passed
all presently performed tests with complete success.  These tests have
probed many features of the structure of general relativity: the metric
nature of the coupling to matter, the slow-motion weak-field limit [which,
in field-theory language, gives already some limits on the field content
of the theory, see eq.~(\ref{eq:5.36})], the effects of strong internal
gravitational fields on the orbital dynamics and the gravitational
radiation reaction of binary systems, and, to a lesser degree, the
large-distance and large-time behaviour of the theory.

Does this mean that one should stop testing Einstein's theory and
consider it as definitely proven~? No, if one remembers that general
relativity has no free parameters (considering that the cosmological
constant belongs to the ``matter" side of Einstein's equations).
Any test of Einstein's theory is a potential killer of the theory.
A clear experimental disproof of Einstein's theory would represent a
major crisis for physics.

It is interesting to discuss whether one can presently think of ways in
which a non-general-relativistic theory would have naturally passed all
existing tests with the same success as Einstein's theory, while still
differing from it in an essential way.  In fact there are several ways
in which this could have happened, and we have already quoted some. One way
relies on the possible existence of short-range contributions to gravity.  For
instance, the kind of (supergravity-motivated) vector partner of $g_{\mu\nu}$
suggested by Scherk \cite{S79}, with gravitational-strength coupling to the
(effective) masses of the quarks and leptons $(g_i = \sqrt{4\pi G} m_i)$ would
have escaped detection so far if its range $\lambda \sim 1$~m.  Other ways use
the fact that the cosmological evolution of the universe at large could
dynamically drive a non-general-relativistic theory to a state where its
predictions are virtually identical to the general relativistic ones
\cite{DN92,40a}. The study of such models can help us in focussing on certain
experiments which are more likely to unravel significant deviations from the
general relativistic predictions.

\bigskip

\noindent{\bf Appendix}

\smallskip

Our signature is $-+++$; we use greek indices to denote spacetime
indices $(\mu,\nu,\cdots =0,1,2,3)$ and latin indices for spatial ones
$(i,j,\cdots =1,2,3)$. The flat (Minkowski) metric is denoted
$f_{\mu\nu} = {\rm diag} (-1,+1,+1,+1) =f^{\mu\nu}$, instead of the
often used $\eta_{\mu\nu}$. To save writing minus signs we define
$g\equiv -\det (g_{\mu\nu})$. When using general vectorial frames
${\bf e}_\alpha$ (not necessarily coordinate ones) we think of the last
lower index on the connection coefficients (i.e. $\delta$ in
$\Gamma^\alpha_{\beta\delta}$) as being the differentiation index:
\[ \nabla_\delta V^\alpha = {\bf e}_\delta (V^\alpha) + \Gamma^\alpha
_{\beta\delta} V^\beta \ , \]
where ${\bf e}_\delta \equiv e^\nu_\delta \partial /\partial x^\nu$ is
the $\delta$-th frame vector $({\bf V}=V^\alpha {\bf e}_\alpha)$ viewed
as a derivative operator. In Cartan language this means using the
connection one-forms $\mbox{\boldmath{$\omega$}}^\alpha_{\ \beta} =
\Gamma^\alpha_{\ \beta\delta} \mbox{\boldmath{$\theta$}}^\delta$ where
$\mbox{\boldmath{$\theta$}}^\alpha$ is the co-frame
($< \mbox{\boldmath{$\theta$}}^\alpha,
{\bf e}_\beta > = \delta^\alpha_\beta$). [See e.g. Ref.~\cite{CBDW}].
Our curvature conventions follow from Cartan's $\Omega^\alpha_{\ \beta}=
d\omega^\alpha_{\ \beta} + \omega^\alpha_{\ \sigma}\
\omega^\sigma_{\ \beta}$, which yields, when using a coordinate basis
$({\bf e}_\mu =\partial/ \partial x^\mu)$,
\[ R^\mu_{\ \nu\rho\sigma} = \partial_\rho \Gamma^\mu_{\ \nu\sigma} +
\Gamma^\mu_{\ \lambda\rho} \Gamma^\lambda_{\ \nu\sigma} -
\{ \rho \leftrightarrow \sigma \} \]
Then
\[ R_{\mu\nu} \equiv R^\lambda_{\ \mu\lambda\nu} = \partial_\lambda
\Gamma^\lambda_{\ \mu\nu} - \cdots \]
and
\[ R \equiv g^{\mu\nu} R_{\mu\nu} \ . \]
With these conventions $R_{\mu\nu}$ and $R$ are positive-definite
for the metrics of spheres.

\end{document}